\documentstyle[psfig]{mn}

\newif\ifAMStwofonts

\ifoldfss
  \ifCUPmtlplainloaded \else
    \NewTextAlphabet{textbfit} {cmbxti10} {}
    \NewTextAlphabet{textbfss} {cmssbx10} {}
    \NewMathAlphabet{mathbfit} {cmbxti10} {} 
    \NewMathAlphabet{mathbfss} {cmssbx10} {} 
  \fi
  \ifAMStwofonts
    \ifCUPmtlplainloaded \else
      \NewSymbolFont{upmath} {eurm10}
      \NewSymbolFont{AMSa} {msam10}
      \NewMathSymbol{\upi}     {0}{upmath}{19}
      \NewMathSymbol{\umu}     {0}{upmath}{16}
      \NewMathSymbol{\upartial}{0}{upmath}{40}
      \NewMathSymbol{\leqslant}{3}{AMSa}{36}
      \NewMathSymbol{\geqslant}{3}{AMSa}{3E}

       \let\le=\leqslant
       \let\ge=\geqslant
    \fi
  \fi
\fi 

\ifnfssone
  \newmathalphabet{\mathit}
  \addtoversion{normal}{\mathit}{cmr}{m}{it}
  \addtoversion{bold}{\mathit}{cmr}{bx}{it}
  \newmathalphabet{\mathbfit} 
  \addtoversion{normal}{\mathbfit}{cmr}{bx}{it}
  \addtoversion{bold}{\mathbfit}{cmr}{bx}{it}
  \newmathalphabet{\mathbfss} 
  \addtoversion{normal}{\mathbfss}{cmss}{bx}{n}
  \addtoversion{bold}{\mathbfss}{cmss}{bx}{n}
  \ifAMStwofonts
    \ifCUPmtlplainloaded \else
      \UseAMStwoboldmath
      \makeatletter
      \new@mathgroup\upmath@group
      \define@mathgroup\mv@normal\upmath@group{eur}{m}{n}
      \define@mathgroup\mv@bold\upmath@group{eur}{b}{n}
      \edef\UPM{\hexnumber\upmath@group}
      \new@mathgroup\amsa@group
      \define@mathgroup\mv@normal\amsa@group{msa}{m}{n}
      \define@mathgroup\mv@bold\amsa@group{msa}{m}{n}
      \edef\AMSa{\hexnumber\amsa@group}
      \makeatother
      \mathchardef\upi="0\UPM19
      \mathchardef\umu="0\UPM16
      \mathchardef\upartial="0\UPM40
      \mathchardef\leqslant="3\AMSa36
      \mathchardef\geqslant="3\AMSa3E

       \let\le=\leqslant
       \let\ge=\geqslant
    \fi
  \fi
\fi 

\ifnfsstwo
  \DeclareMathAlphabet{\mathbfit}{OT1}{cmr}{bx}{it}
  \SetMathAlphabet\mathbfit{bold}{OT1}{cmr}{bx}{it}
  \DeclareMathAlphabet{\mathbfss}{OT1}{cmss}{bx}{n}
  \SetMathAlphabet\mathbfss{bold}{OT1}{cmss}{bx}{n}
  \ifAMStwofonts
    \ifCUPmtlplainloaded \else
      \DeclareSymbolFont{UPM}{U}{eur}{m}{n}
      \SetSymbolFont{UPM}{bold}{U}{eur}{b}{n}
      \DeclareSymbolFont{AMSa}{U}{msa}{m}{n}
      \DeclareMathSymbol{\upi}{0}{UPM}{"19}
      \DeclareMathSymbol{\umu}{0}{UPM}{"16}
      \DeclareMathSymbol{\upartial}{0}{UPM}{"40}
      \DeclareMathSymbol{\leqslant}{3}{AMSa}{"36}
      \DeclareMathSymbol{\geqslant}{3}{AMSa}{"3E}

       \let\le=\leqslant
       \let\ge=\geqslant
    \fi
  \fi
\fi 

\ifCUPmtlplainloaded \else
  \ifAMStwofonts \else 
    \def\upi{\pi}
    \def\umu{\mu}
    \def\upartial{\partial}
  \fi
\fi

\title[Constraints on the cosmic matter density]
       {The ROSAT-ESO Flux-Limited X-Ray (REFLEX) Galaxy Cluster
         Survey\,VI: Constraints on the cosmic matter density from the
         KL power spectrum}
\author[P. Schuecker et al.]
       {Peter Schuecker$^1$, Luigi Guzzo$^2$,
        Chris A. Collins$^3$ and Hans B\"ohringer$^1$ \\
    $^{1}$ Max-Planck-Institut f\"ur extraterrestrische Physik,
             Giessenbachstra{\ss}e 1, 85740 Garching, Germany\\ 
    $^{2}$ Osservatorio Astronomico di Brera, via Bianchi, 22055
Merate (LC), Italy\\
    $^{3}$ Astrophysics Research Institute, Liverpool John Moores
University, Birkenhead CH41 1LD,
Great Britain\\
}

\date{Accepted 20\_\_ \_\_ \_\_. Received 20\_\_ \_\_ \_\_}
\pagerange{\pageref{firstpage}--\pageref{lastpage}}
\pubyear{2002}

\begin{document}

\maketitle

\label{firstpage}

\begin{abstract}
The Karhunen-Lo\'{e}ve (KL) eigenvectors and eigenvalues of the sample
correlation matrix are used to analyse the spatial fluctuations of the
REFLEX clusters of galaxies. The method avoids the disturbing effects
of correlated power spectral densities which affects all previous
cluster measurements on Gpc scales. Comprehensive tests use a large
set of independent REFLEX-like mock cluster samples extracted from the
Hubble Volume Simulation. It is found that unbiased measurements on
Gpc scales are possible with the REFLEX data. The distribution of the
KL eigenvalues are consistent with a Gaussian random field on the
93.4\% confidence level. Assuming spatially flat cold dark matter
models, the marginalization of the likelihood contours over different
sample volumes, fiducial cosmologies, mass/X-ray luminosity relations
and baryon densities, yields the 95.4\% confidence interval for the
matter density of $0.03<\Omega_mh^2<0.19$. The N-body simulations show
that cosmic variance, although difficult to estimate, is expected to
increase the confidence intervals by about 50\%.
\end{abstract}


\section{Introduction}\label{INTRO}

The cosmological parameters characterize the time evolution of the
cosmic scale factor, and determine the formation and evolution of
structures within the Universe. Rich clusters of galaxies are
physically well-defined tracers of these structures because they can
only be formed at well-defined sites, namely where the peaks of the
initial density field exceed a critical density threshold. This
threshold is soley determined by gravitation. Gaussian initial
conditions simplify the situation even more. Therefore, the physical
properties of the cluster population, like mass function and spatial
distribution, are closely related to the global properties of the
Universe and give thus direct information on the values of the
cosmological parameters.

Important constraints on the values of the cosmological parameters
obtained with galaxy clusters are generally based on measurements of
the mean cluster abundance (e.g., Viana \& Liddle 1996, Bahcall \& Fan
1998, Borgani et al. 2001, Reiprich \& B\"ohringer 2002, see also the
theoretical work of Haiman, Mohr \& Holder 2001). However, the cluster
abundance probes only a small scale range so that the resulting values
of the matter density and the normalization parameter, $\sigma_8$, of
the structure formation models are highly correlated.

Measurements of the spatial fluctuations of the cluster abundance over
a sufficiently large scale range can break the degeneracy. A review of
recent measurements obtained with the spatial two-point correlation
function of galaxy clusters is given in Collins et al. (2000). The
fluctuations are also characterised by the power spectrum, $P(k)$,
which is directly related to theory. Recent measurements of this
quantity use either optically selected clusters (Peacock \& West 1992,
Einasto et al. 1993, Jing \& Valdarnini 1993, Einasto et al. 1998,
Retzlaff et al. 1998, Tadros et al. 1998, Miller \& Batuski 2001) or
X-ray selected clusters (Retzlaff 1999, Schuecker et al. 2001,
Zandivarez, Abadi \& Lambas 2001). The advantages of X-ray against
optically selected cluster samples are discussed in, e.g., Borgani \&
Guzzo (2001).

For the construction of the ROSAT ESO Flux Limited X-Ray (REFLEX)
cluster sample special care was taken to get a homogeneous sampling
and a high completeness (B\"ohringer et al. 2001). The sample consists
of 452 clusters with redshifts $z\le 0.45$, selected in X-rays from the
ROSAT All-Sky Survey and is confirmed by extensive optical follow-up
observations within a large ESO Key Programme (B\"ohringer et
al. 1998, Guzzo et al. 1999). This makes the sample well-suited for
spatial analyses on Gpc scales. 

However, on Gpc scales the anisotropy of the volumes of all cluster
surveys becomes apparent. Therefore, reliable $P(k)$ measurements of
the projects mentioned above could only be obtained up to maximum
scales reaching 200 to $400\,h^{-1}\,{\rm Mpc}$. Unfortunately, the
plane waves used in the standard power spectrum analyses to expand the
observed fluctuations are no longer orthogonal on Gpc scales and must
be replaced by another set of basis functions fulfilling this
fundamental criterion. The conditions of orthogonality of the basis
functions and statistical orthogonality of the expansion coefficients
lead to the Karhunen-Lo\'{e}ve (KL) eigenvectors of the sample
correlation matrix (Karhunen 1947, Lo\'{e}ve 1948). They offer an
analysis of the cluster power spectrum which is free from any
disturbing effects of correlated power spectral densities affecting
all previous cluster measurements on Gpc scales.

The present paper applies the KL method to estimate the cosmic matter
density and the linear normalization of the matter power spectrum
using the spatial fluctuations of the REFLEX clusters. In order to
introduce the basic quantities and to make the paper more
self-contained, we recall in Sect.\,\ref{EIGEN} some aspects of the KL
method and its application to large-scale structure work. The
relations between the observed quantities as measured in the present
investigation and the cosmological parameters are derived in
Sect.\,\ref{MODEL}. The basic properties of the REFLEX cluster sample
are summarized in Sect.\,\ref{REFLEX}. The KL eigenvectors and the
spectrum of the eigenvalues of the REFLEX sample are presented in
Sect.\,\ref{EIGENMODES}. The final results on the cosmic matter
density and normalization of the matter power spectrum obtained with
the REFLEX sample are given in Sect.\,\ref{RESULTS} and are discussed
in Sect.\,\ref{DISCUSS}.

To evaluate systematic and statistical errors as well as the effects
of cosmic variance, end-to-end tests are performed which follow the
basic steps of the REFLEX survey reduction and the KL method of
parameter estimation. Here we use a large set of independent
REFLEX-like mock cluster samples selected from the Hubble Volume
Simulation. The details are given in Appendix \ref{TESTS}.

As the fiducial cosmological model which is used to compute geometric
quantities and KL eigenvectors, we assume a pressure-less, spatially
flat Friedmann-Lema\^{\i}\-tre model, the cosmic matter density,
$\Omega_m=0.3$, the cosmological constant in the form
$\Omega_\Lambda=0.7$, and the Hubble constant in units of
$h=H_0/100\,{\rm km}\,{\rm s}^{-1}\,{\rm Mpc}^{-1}$.

\section{The KL method}\label{EIGEN}

The KL method was first used to test cosmological structure formation
models by Bond (1995) using cosmic microwave background (CMB)
temperature maps. Vogeley \& Szalay (1996) translated the method to
the case of the spatial analysis of galaxy distributions. Applications
to galaxy surveys can be found in Matsubara, Szalay \& Landy (2000)
and Szalay et al. (2001). The KL method as used here to analyse
cluster data consists of two steps: calculation of the eigenvectors
(Sect.\,\ref{EIGEN_KL}), and likelihood estimation of the values of
the power spectrum (cosmological) parameters which maximizes the
probability of the observed fluctuations (Sect.\,\ref{EIGEN_MOD}).

\subsection{Calculation of the eigenvectors}\label{EIGEN_KL} 

The survey volume is devided into $M$ cells, each with a specific
comoving volume, $V_i$. We chose spherical coordinates and specified
each edge of a cell by the three normal Euler coordinates. The results
of the KL analysis do, however, not depend on a specific pixellation
(see below).

In the $i$-th cell centered on the comoving coordinate vector
$\vec{r}_i$, $D_i$ clusters are counted. The expansion of the field,
$D_i$, can be written in the component form as
$D_i\,=\,\sum_{j=1}^M\,\psi_{ij}\,B_j\,,\quad i=1,\ldots,M$, where the
$\psi_{ij}$ are the elements of a matrix which gives the $i$-th
component of the $j$-th basis vector.

The modes and coefficients should fulfill two criteria. (i) The modes
should be orthogonal to each other, $\sum_{k=1}^M\,\psi_{ik}^{\rm
T}\,\psi_{kj}\,=\,\delta_{ij}$, where T denotes the transpose of a
matrix and $\delta_{ij}$ the Kronecker delta.  (ii) The modes should
yield statistically orthogonal expansion coefficients. One thus
requires that the expectation value of the sample covariance matrix
has the form $<B_i\,B_j^{\rm T}>\,=\,<B_j^2>\,\delta_{ij}$. The two
criteria directly lead to the equations which determine the optimal
basis vectors,
\begin{equation}\label{KL7}
\sum_{l=1}^MR_{kl}\,\psi_{jl}\,=\,<B_j^2>\,\psi_{kj}\,=\,\lambda_j\,\psi_{kj}\,,
\end{equation}
with the components of the correlation matrix, $R_{kl}$, defined via
the expectation values, $<B_i\,B_j^{\rm
T}>\,=\,\sum_{k,l=1}^M\,\psi_{ik}^{\rm T}\,R_{kl}\,\psi_{lj}$, through
$R_{kl}\,=\,<D_k\,D_l^{\rm T}>$.  The problem of finding the set of
modes satisfying the conditions of orthogonality and statistical
orthogonality thus reduces to the problem of finding the eigenvectors
of the correlation matrix $R$, called the KL eigenvectors, and the
corresponding eigenvalues, $\lambda_i=<B_i^2>$, constituting the KL
fluctuation spectrum.

For an arbitrary pixellation of the survey volume the noise per
counting cell varies even for volume-limited samples, and one has to
diagonalize the noise component of the correlation matrix before the
eigenvectors are computed. The separation of signal and noise in the
new basis is achieved by transforming (whitening) the elements of the
correlation matrix computing
$R'_{ij}\,=\,\sum_{k,l=1}^MN^{-1/2}_{ik}\,R_{kl}\,N^{-1/2}_{lj}$. The
$N^{-1/2}_{ik}$ are the inverse square roots of the elements of the
noise correlation matrix,
$N_{ik}\,=\delta_{ik}\,\int_{V_{i}}<n(\vec{r})>\,d^3r\,=\,N_i$, and
$<n(\vec{r})>$ the expected cluster number density at the comoving
position $\vec{r}$.

\subsection{Estimation of model parameters of the power
spectrum}\label{EIGEN_MOD} 

The present investigation tests the fluctuating part of the cluster
number counts. Therefore, the covariance matrix, $C$, of the KL
coefficients is used to estimate the values of the (cosmological)
parameters, $x_1,\ldots,x_q$, characterizing the power spectrum.

The covariance matrix is estimated in the following way. Choose a
specific set of $x_i$ values to specify the model
$P(k)$. Fourier-transform $P(k)$ by direct numerical integration in
order to get the correlation function, $\xi(r)$, and compute the
continuous part of the cell-averaged correlation matrix (we use a
Monte-Carlo estimate) of the model to be tested:
\begin{equation}\label{XI1}
\xi_{ij}\,=\,\frac{1}{V_i\,V_j}\,
\int_{V_i}d^3\vec{r}_i\,\int_{V_j}d^3\vec{r}_j\,\xi(|\vec{r}_i-\vec{r}_j|)\,.
\end{equation}
Choose also an appropriate model for the expected average number of
clusters, $N_i$, in each cell. To be consistent with the fluctuation
analysis we use an empirical model (see Sect.\,\ref{MEAN}). This model
is not changed during the testing of different model power
spectra. The coefficients
\begin{equation}\label{ML_3}
C_{ij}=\sum_{k,l=1}^M \frac{\psi_{ik}^{\rm
T}}{\sqrt{N_k}}\,(N_k\,N_l\,\xi_{kl}\,+\,N_k\,\delta_{kl})\,
\frac{\psi_{lj}}{\sqrt{N_l}}
\end{equation}
constitute the estimated model covariance matrix, $C$, of the KL
coefficients, where the first term on the right-hand side of
(\ref{ML_3}) describes the clustering signal and the second term the
noise. The $\psi_{ij}$ are the KL eigenvectors of the whitened
correlation matrix obtained with the fiducial cosmology (see
Sect.\,\ref{EIGEN_KL}). The model covariance matrix is not diagonal
unless the fiducial model used to compute the KL eigenvectors is
identical to the model used to compute the $\xi_{ij}$.

In Sect.\,\ref{EIGENMODES} it will be shown that the frequency
distribution of the REFLEX KL coefficients, $B_i$, is well described
by a Gaussian. Due to the linearity of the KL transform this suggests
that the REFLEX cluster density field is governed by a Gaussian-like
random field (for large cell sizes). The multivariate likelihood
function of the parameters $x_i$ should thus be of the form
\begin{eqnarray}\label{ML_4}
{\cal L}(B_1,\ldots,B_M|x_1,\ldots,x_q)\,=\nonumber\\
      (2\pi)^{-M/2}|{\rm det}\,C|^{-1/2}
      \exp\left(\,-\frac{1}{2}\Delta\vec{B}^T\,\,
      C^{-1}\,\,\Delta\vec{B}\,\right)\,,
\end{eqnarray}
with the difference vector $\Delta\vec{B}=\vec{B}-<\vec{B}>$. The
$x_i$ values of the power spectrum parameters which maximise the
probability of obtaining fluctuations transformed into the KL base as
large as observed are defined by the maximum of the sample function
(\ref{ML_4}).

\begin{figure*}
\vspace{-1.0cm}
\centerline{\hspace{-11.5cm}
\psfig{figure=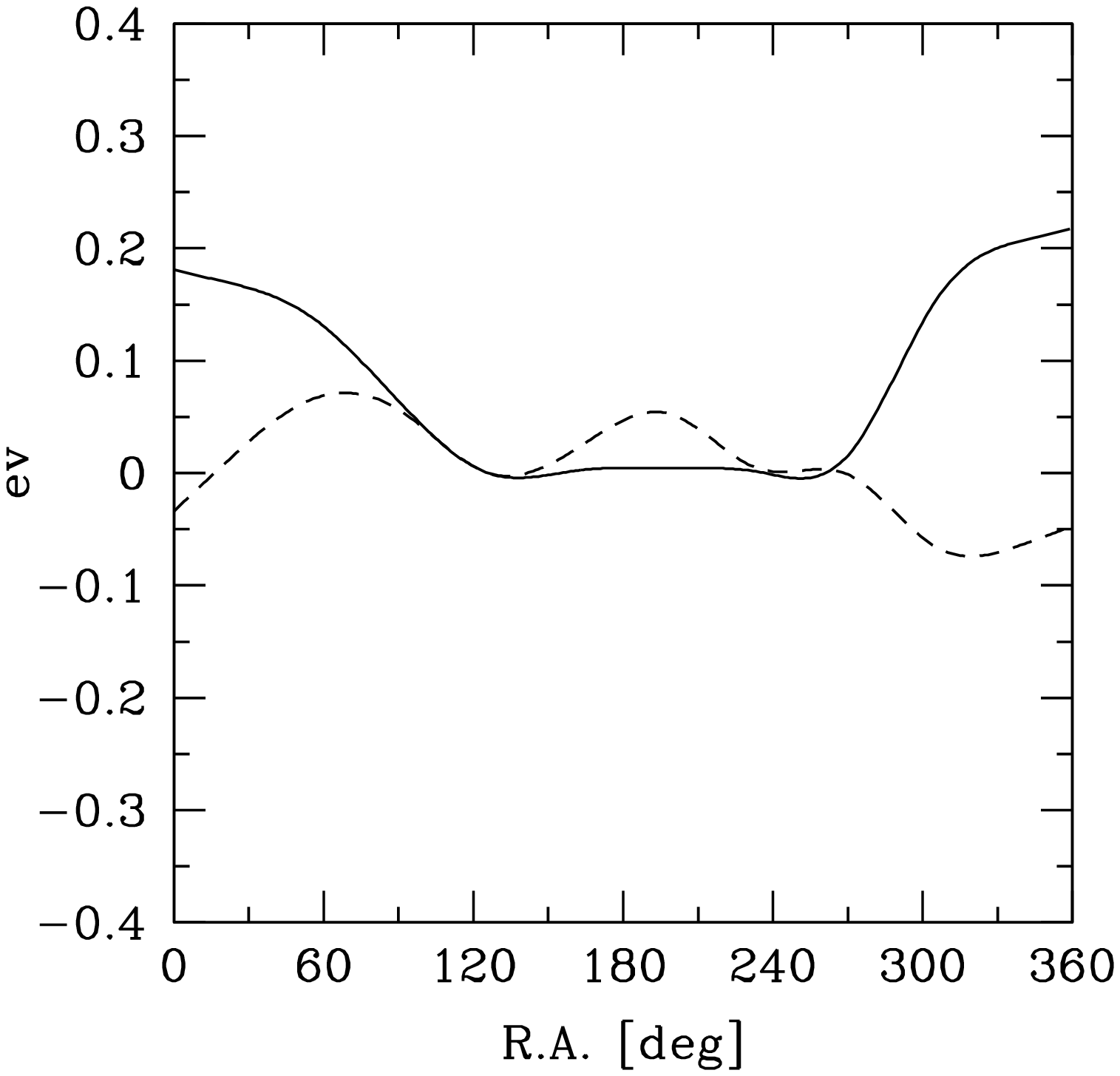,height=7.5cm,width=7.5cm}}
\vspace{-7.5cm}
\centerline{\hspace{0.25cm}
\psfig{figure=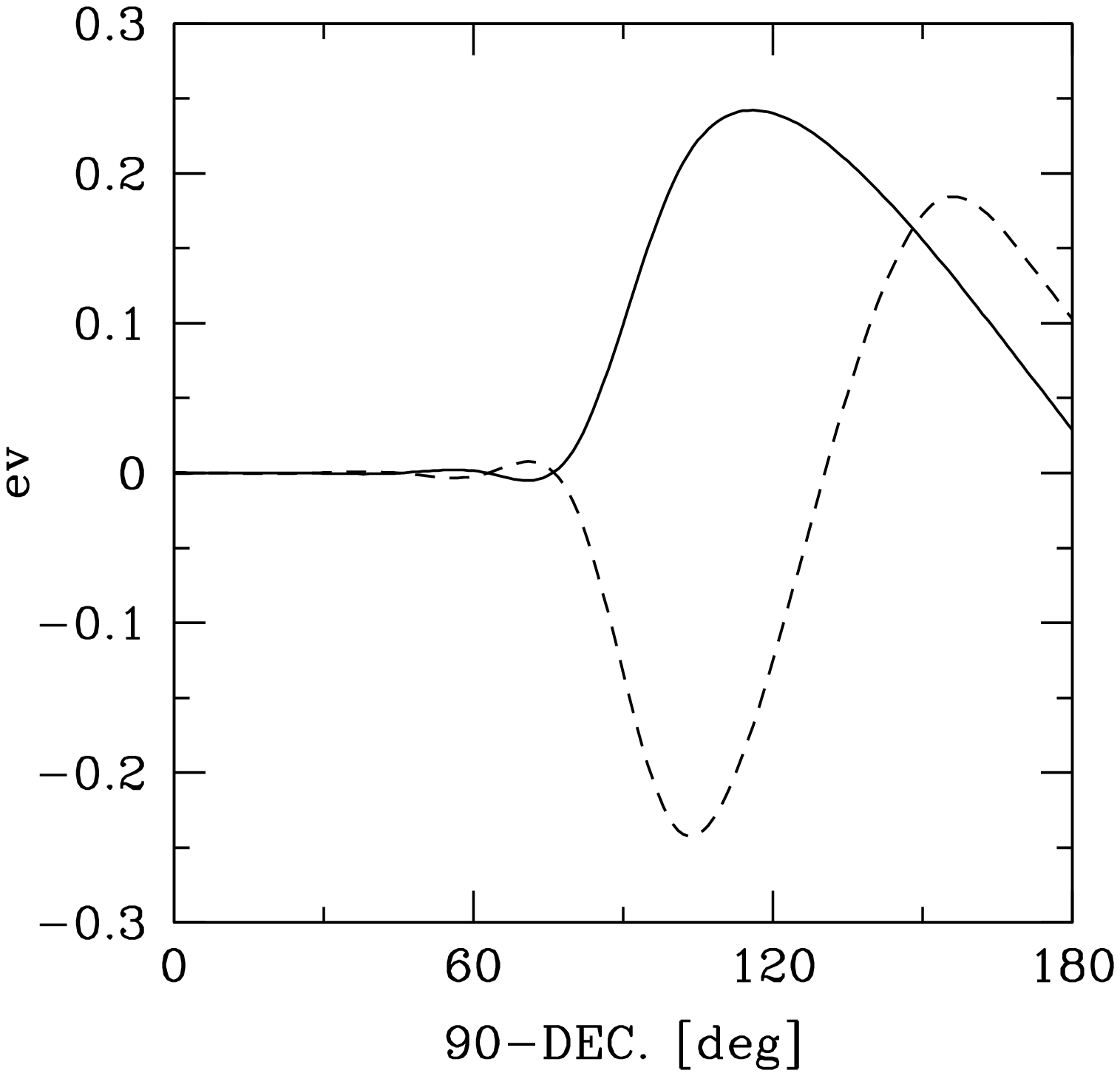,height=7.5cm,width=7.5cm}}
\vspace{-7.5cm}
\centerline{\hspace{11.5cm}
\psfig{figure=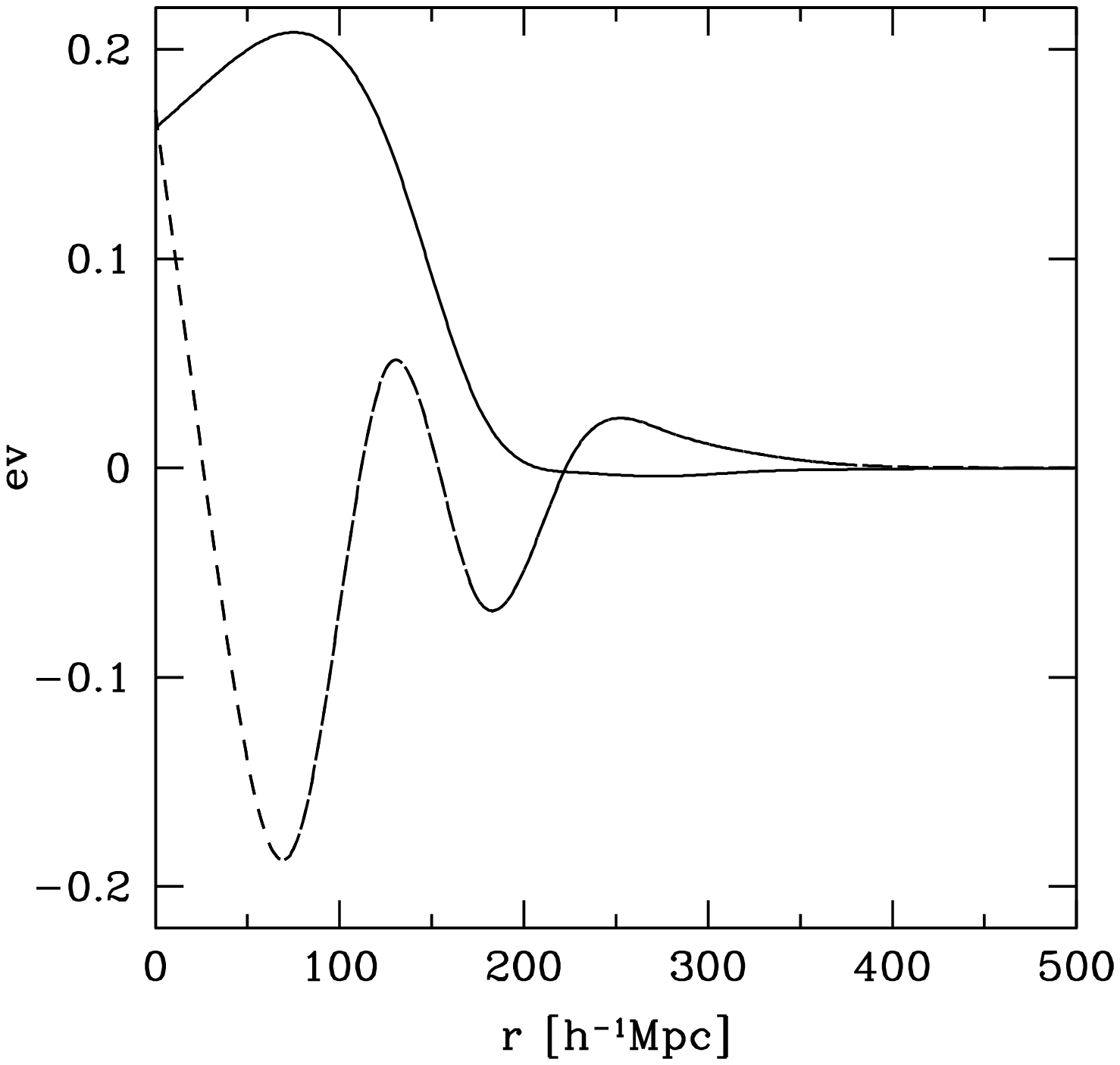,height=7.5cm,width=7.5cm}}
\vspace{-1.0cm}
\caption{\small Examples of one-dimensional tracings of the
three-dimensional KL eigenvectors as a function of Right Ascension
(left), Declination (middle), and comoving radial distance
(right). The values of the eigenvectors are computed for each
direction (in arbitrary physical units) at the centers the 10 cells and
are then interpolated (for illustration) by cubic splines. Continuous
lines show eigenvectors with the highest eigenvalue, dashed lines with
lower eigenvalues (higher orders). The REFLEX survey crosses the
galactic plane at ${\rm R.A.}=120$\,deg and 270\,deg, and has no
clusters in the North as seen by the low values of the eigenvectors at
the corresponding positions. The effective depth of the REFLEX survey
is at $r=150\,h^{-1}\,{\rm Mpc}$.}
\label{FIG_EV}
\end{figure*}

\section{The relations between the observed quantities and 
the cosmological parameters}\label{MODEL}

The KL method estimates the values of the cosmological parameters
comparing the observed fluctuations of the cluster number densities
transformed into the KL eigenvector basis with theoretical
expectations. In the following the general assumptions on the matter
power spectrum (see Sect.\,\ref{TRANS}), on the relation between the
observed amplitude of the cluster power spectrum and the standard
$\sigma_8$ normalization of the matter power spectrum (see
Sect.\,\ref{RELATION}), and on the empirical model used to compute the
expected mean cluster number counts, $N_i$, (see Sect.\,\ref{MEAN})
are described.

\subsection{Matter power spectrum}\label{TRANS}

On the largest scales the density field is assumed to be Gaussian with
a power-law spectrum of adiabatic matter fluctuations, $P(k)\sim
k^n$. In order to describe the matter power spectrum on smaller
scales, we take into account the effects of a collisionless matter
component and the collisional baryonic component.  Instead of solving
the corresponding multispecies Boltzmann equations for each model to
be tested, the comparatively simple fitting formulae for the transfer
functions, $T_x(k)$, given in Eisenstein \& Hu (1998) are used
providing a more accurate description than the standard BBKS fitting
functions mainly characterized by the scale-independent shape
parameter, $\Gamma$.

We restrict the present KL analyses to the estimation of the matter
density, $\Omega_m=\vec{x}$ and $\sigma_8$ because they determine --
for a given Hubble constant -- the general shape and amplitude of the
power spectrum. For the Hubble constant we take $h=0.7$ as suggested
by Hubble Space Telescope observations (Freedman et al. 2001). The
final results on $\Omega_m$ are given in units of $h$. In addition, we
assume $n=1.0$, a mean temperature of the CMB of $T_{\rm CMB}=2.728$,
a spatially flat cosmology as suggested by CMB measurements (De
Bernardis et al. 2000), and the baryon density $\Omega_bh^2=0.0196$ as
suggested by chemical abundance measurements of distant quasars
(Burles \& Tytler 1998) and Standard Big Bang Nucleosynthesis
calculations (Burles, Nollett \& Turner 2001). In the next paper
additional observational constraints will be included so that the
priors can be weakened.

\subsection{Relation between $\sigma_8$ and the observed
amplitude, $P_0$, of the power spectrum}\label{RELATION}

It is generally assumed that on large scales structure growth can be
treated within linear theory. The observed cluster power spectrum,
$P_{\rm obs}(k)$, is the result of a complex averaging process over
evolving matter power spectra, $P(k,z)=P(k)D^2(z)/D^2(0)$ and clusters
with different values of the biasing parameter, $b(M,z)$. Here, $D(z)$
is the linear growth factor. Matarrese et al. (1997) and Moscardini et
al. (2000) derived analytic equations approximating this process which
we summarize by the equation
\begin{equation}\label{A0}
P_{\rm
obs}(k)\,=\,\left<P(k,z)\,\,\left<b(M,z)\right>^2_M\right>_Z\,,
\end{equation}
where the mass and redshift expectations involve the actual number of
clusters, $N(M,z)dMdz$, observed within given mass and redshift
shells, and the corresponding redshift histogram, $N(z)dz$. Here,
$<b(M,z)>_M$ is the mass-weighted biasing factor and $<\cdot>_Z$ the
redshift average (eq.\,14 in Moscardini et al. 2000). Within the
general framework of linear perturbation theory of cosmic structures,
the present-day matter power spectrum is
\begin{equation}\label{A3}
P(k)=\frac{2\pi^2\,\sigma^2_8\,k^n\,T^2_x(k)}{\int\,dk\,
k^{n+2}\,T^2_x(k)\,|W(8k/h{\rm
Mpc}^{-1})|^2}\,,
\end{equation}
or
\begin{equation}\label{A31}
P(k)\,=\,\sigma^2_8\,\,k^n\,\,T^2_x(k)\,\,\zeta^{-1}_{n\,x\,8}\,,
\end{equation}
where we have introduced for convenience the function
\begin{equation}\label{A5}
\zeta_{n\,x\,R}\,=\,\frac{1}{2\pi^2}\,\int\,dk\,k^{n+2}\,T^2_x(k)\,|W(kR)|^2\,.
\end{equation}
The spectrum is normalized by $\sigma_8$ in the standard way using the
Fourier-transformed top-hat filter, $W(kR)$, with the comoving radius
$R=8\,h^{-1}\,{\rm Mpc}$. It is important to note that $\sigma_8$
defined in this way reflects the amplitude of the power spectrum
without any non-linear corrections. 

Equation\,(\ref{A0}) implies that the observed mass and
redshift-averaged cluster power spectrum has the same shape, i.e.,
functional form as the underlying matter power spectrum,
$P(k)$. Therefore, we set $P_{\rm obs}(k)=P_0k^nT^2_x(k)$ with the
parameters $\vec{x}$ and $P_0$ determined by observation. Equating the
latter formula and the theoretical expectation (\ref{A0}), yields the
general relation between the observed amplitude and the normalization
of the matter power spectrum,
\begin{equation}\label{A4}
P_0\,=\,\sigma^2_8\,\zeta^{-1}_{n\,x\,8}\,
\left<\frac{D^2(z)}{D^2(0)}\left<b(M,z)\right>_M^2\right>_Z\,.
\end{equation}
As expected, the observed amplitude, $P_0$, of the power spectrum
depends on the sample. Note that in the present case, the actual
values of the cluster sample are inserted in (\ref{A4}), where the
masses are obtained from the observed X-ray luminosities using the
empirical mass-to-X-ray luminosity relation (see eq.\,\ref{ML}) of
Reiprich \& B\"ohringer (2002).

Equation\,(\ref{A4}) shows that a specific biasing model has to be
chosen in order to deduce from $P_0$ the linear normalization,
$\sigma_8$. 

\subsubsection{High-peak biasing}\label{HP_BIAS}

The KL likelihood analysis is based on the assumption of a Gaussian
random field, supported by the observed distribution of the REFLEX KL
eigenvalues (see Sect.\,\ref{EIGENMODES}). In the line of this
observation we apply the related biasing scheme,
$b(M,z)=\delta_c(z)/\sigma^2(M,z)$, derived by Kaiser (1984) for
galaxy clusters on the same statistical grounds as a Gaussian random
field in the limit of high density peaks. The two conditions,
$\sigma(M,z)\ll \delta_c(z)$, and
$(\delta_c(z)/\sigma^2(M,z))^2\xi(r,z)\ll 1$, are generally fulfilled
in the present case because the fluctuation analyses are performed
with massive clusters where $1<b<4$ on $50<r<1000\,h^{-1}\,{\rm Mpc}$
scales where the matter correlation function is about
$10^{-2}>|\xi|>10^{-6}$. The results obtained with the simulations
shown in Appendix\,\ref{TESTS} are also consistent with these
assumptions. In the Kaiser model the biasing parameter is determined
by the slightly redshift-dependent critical density threshold,
$\delta_c(z)$, of the spherical collapse model, and the mass variance,
$\sigma^2(M,z)$. The latter quantity can be written in terms of
$\zeta$ as
\begin{equation}\label{A7}
\sigma^2(M(R),z)\,=\,\sigma^2_8\,\,\frac{D^2(z)}{D^2(0)}\,\,
\frac{\zeta_{n\,x\,R}}{\zeta_{n\,x\,8}}\,.
\end{equation}
The variance in eq.\,(\ref{A7}) decreases with $z$ in a manner that at
high redshift the biasing for clusters of a given mass is stronger
than the decreasing matter power spectrum. For galaxy clusters, $P_0$
is thus expected to increase with $z$. Independent from any
redshift-dependent effect, the high-peak biasing gives the monotonic
relation $P_0\sim 1/\sigma_8^2$.

\subsubsection{Other biasing schemes}\label{O_BIAS}

Based on the Press-Schechter prescription, Mo \& White (1996) derived
a formula which describes the biasing of galaxy-size objects,
\begin{equation}\label{A6}
b(M,z)\,=\,1\,+\,\frac{\delta_c(z)}{\sigma^2(M,z)}\,-\,\frac{1}{\delta_c(z)}\,.
\end{equation}
As for the high-peak biasing, the Mo \& White biasing depends via the
second term on the right-hand side of (\ref{A6}) on $\sigma_8$,
however, now with two additional terms. The first term describes the
peculiar motions of the dark matter haloes and the second and third
terms the effects of the peak-background split. In contrast to the
high-peak biasing, the relation between $\sigma_8^2$ and $P_0$ has a
quadratic form. Therefore, the model suggests two values of $\sigma_8$
for a given $P_0$. The high $\sigma_8$ case characterizes an almost
unbiased halo distribution where basically each mass peak corresponds
to a virialized object (low-biasing regime, unrealistic case for
clusters).  The small $\sigma_8$ case characterizes a strongly biased
distribution where the virialized structures appear as rare objects
(high biasing regime, realistic case for clusters).

The biasing formula given in Sheth \& Tormen (1999) has the same
properties as (\ref{A6}). It is, however, better calibrated with
N-body simulations over a mass range reaching
$5\,10^{13}\,h^{-1}\,M_\odot$ (or the X-ray luminosity
$6\,10^{42}\,h^{-2}\,{\rm erg}\,{\rm s}^{-1}$ for the energy range
$0.1-2.4$\,keV using eq.\,\ref{ML}). Unfortunately, this maximum X-ray
luminosity is close to the minimum X-ray luminosity for completeness
of the cluster sample used in the present investigation (see
Sect.\,\ref{REFLEX}).

\begin{figure}
\vspace{-0.5cm}
\centerline{\hspace{-0.5cm}
\psfig{figure=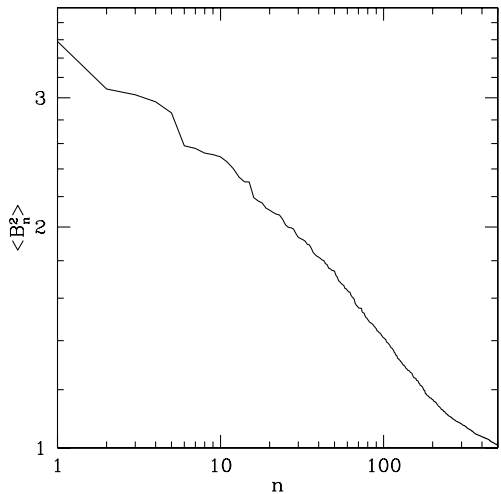,height=8.0cm,width=8.0cm}}
\vspace{-0.75cm}
\caption{\small The spectrum of the KL eigenvalues, $<B_n^2>$, as the
function of rank of the REFLEX cluster sample obtained for the
fiducial cosmology.}
\label{FIG_EW_RANK}
\end{figure}

\begin{figure}
\vspace{-0.5cm}
\centerline{\hspace{-0.5cm}
\psfig{figure=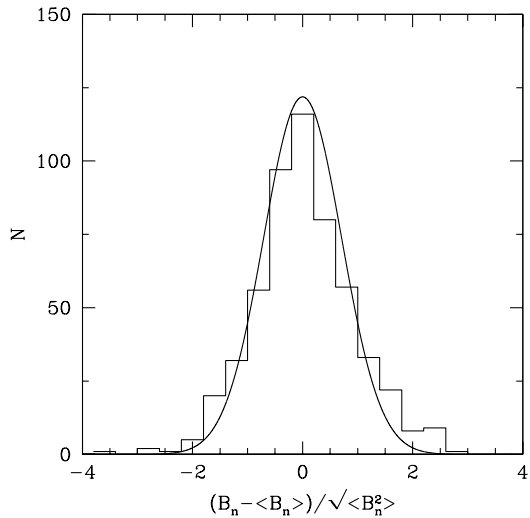,height=8.0cm,width=8.0cm}}
\vspace{-0.75cm}
\caption{\small Histogram of the normalized KL eigenvalues (mean
0.059, standard deviation 0.968) with superposed normal Gaussian
profile for the REFLEX cluster sample.}
\label{FIG_DB_N}
\end{figure}

\subsection{Empirical model for the average cluster number
densities}\label{MEAN}

The present investigation concentrates on the exact modelling of the
fluctuating part of the cluster number counts. The model for the
average cluster number densities, $N_i$ (as used in eq.\,\ref{ML_3}),
is thus not changed during the likelihood optimization. The $N_i$ are
estimated by the following empirical Monte-Carlo method.

For an X-ray cluster sample, the angular part of $N_i$ is mainly
determined by the local X-ray flux limit of the survey, which in turn
is given by the preset nominal flux limit of the sample (see
Sect.\,\ref{REFLEX}), the minimum number of source counts required for
a safe detection, the local satellite's exposure time, and the local
galactic neutral hydrogen column density, $N_{\rm HI}$ (see
Sect.\,\ref{REFLEX} and B\"ohringer et al. 2001). Random number
distributions are computed to generate angular distributions which
precisely follow the survey boundaries as described in Collins et
al. (2000) and Schuecker et al. (2001).

For the radial part of $N_i$ we also generate random distributions
which are now guided by the observed redshift histogram smoothed with
a Gaussian filter profile with the standard deviation
$\sigma_z=0.015$. We compared the KL likelihood contours obtained with
the smoothing method and with the X-ray luminosity function given in
B\"ohringer et al. (2002). No significant differences are found in the
values of the estimated parameters as long as a significant number of
clusters have comoving distances reaching $\ge 300\,h^{-1}\,{\rm
Mpc}$.

\begin{figure*}
\vspace{-0.0cm}
\centerline{\hspace{-10.0cm}
\psfig{figure=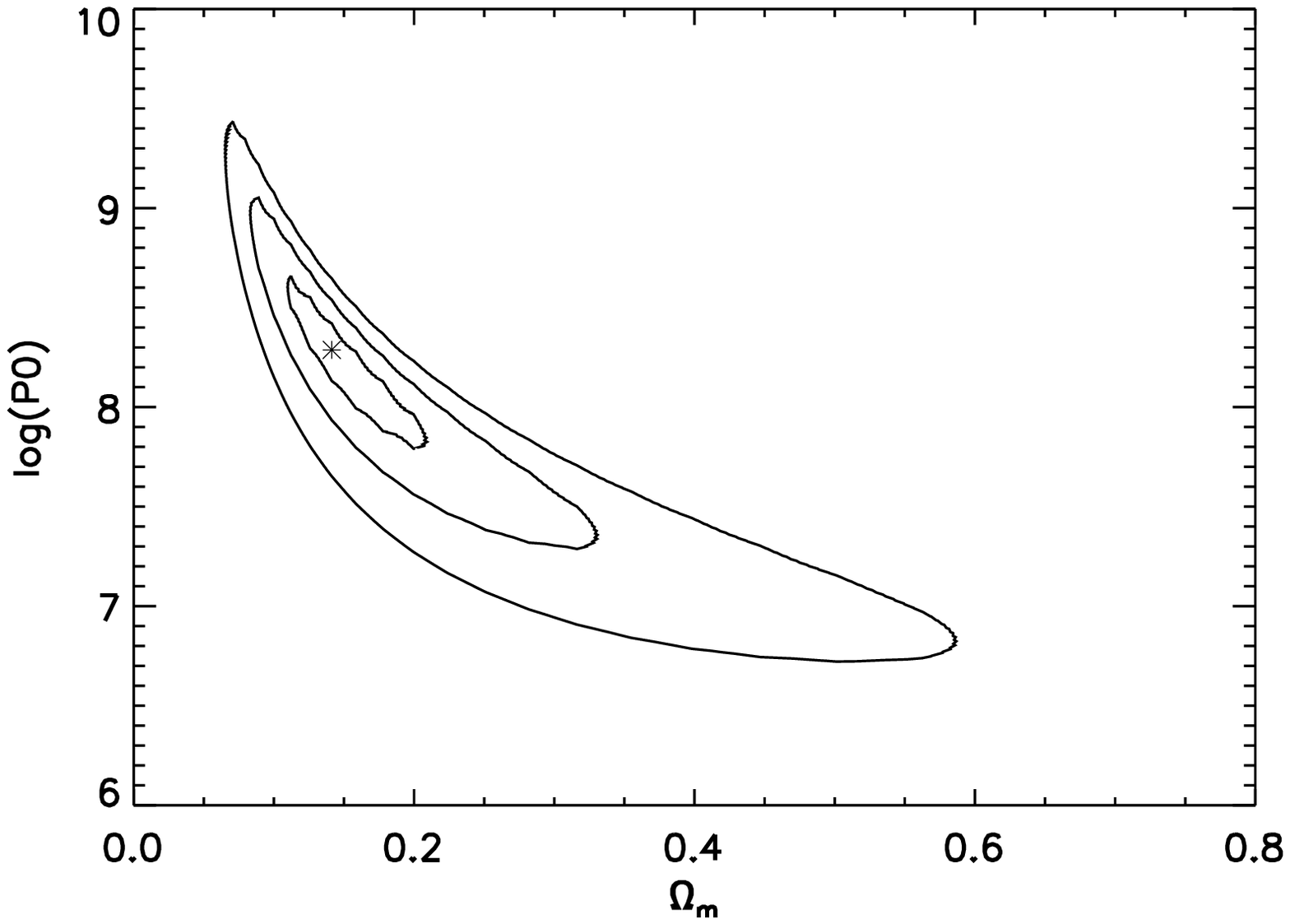,height=5.5cm,width=9.0cm}}
\vspace{-5.5cm}
\centerline{\hspace{ 8.2cm}
\psfig{figure=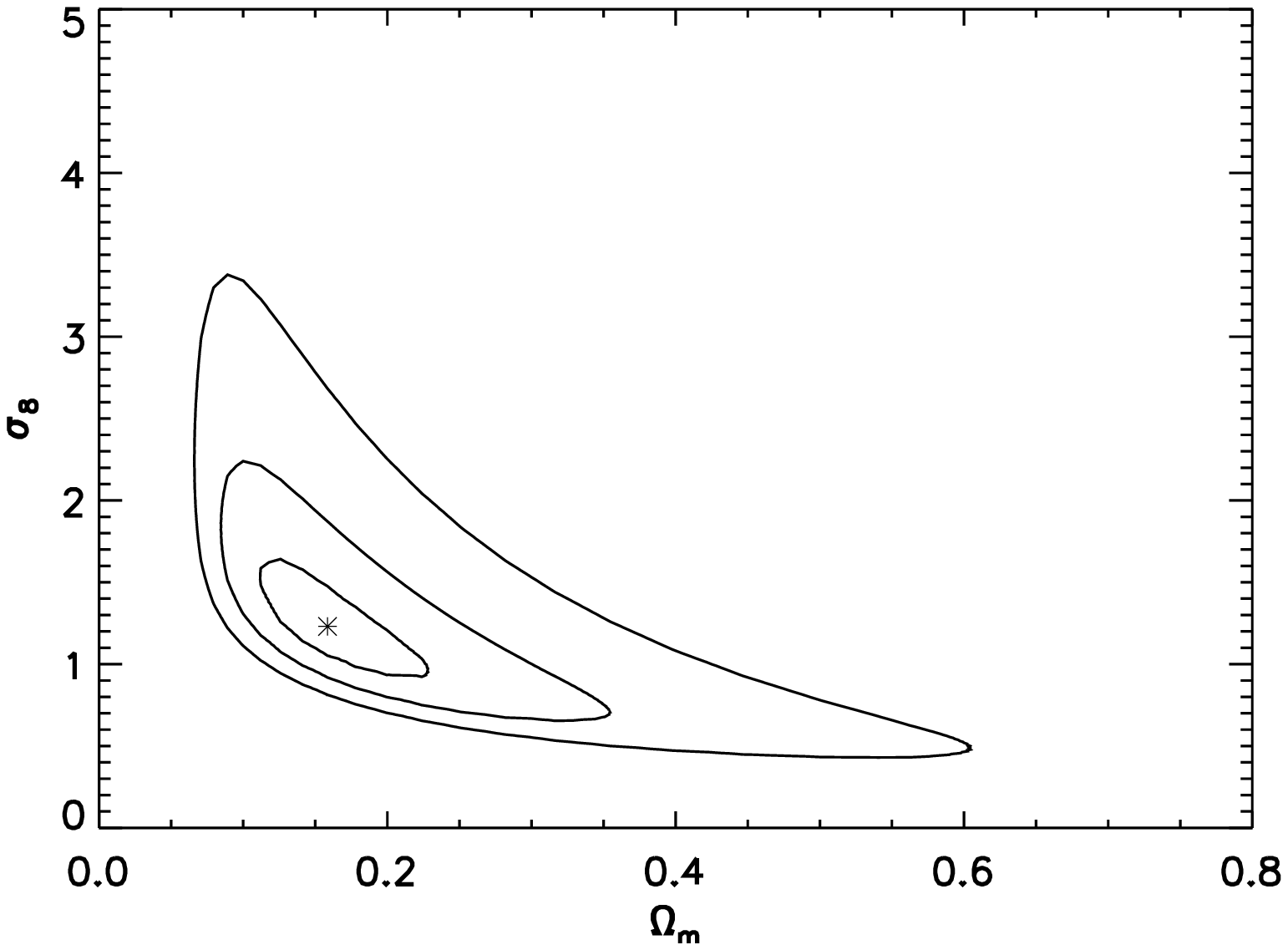,height=5.5cm,width=9.0cm}}
\vspace{-0.5cm}
\centerline{\hspace{-10.0cm}
\psfig{figure=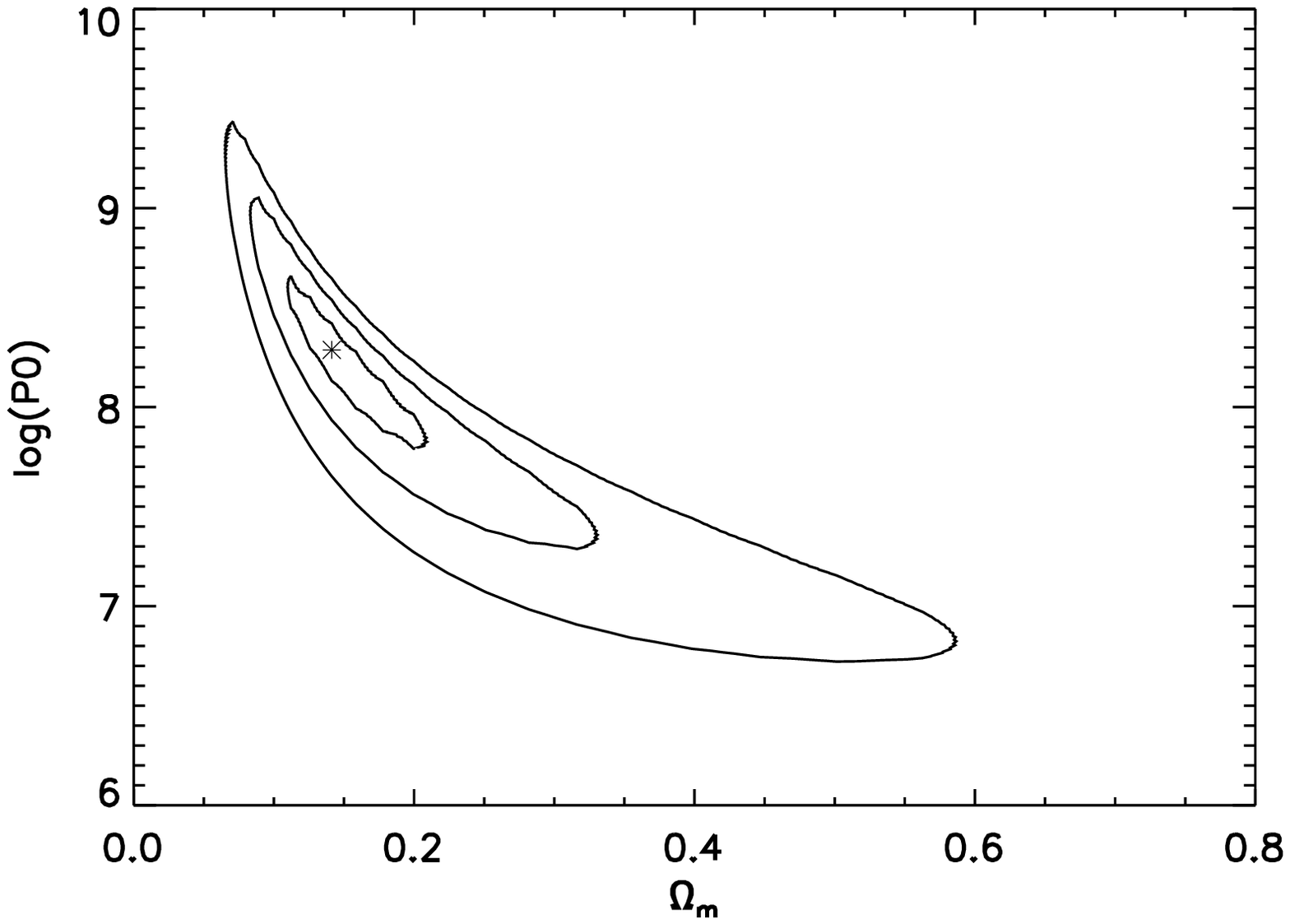,height=5.5cm,width=9.0cm}}
\vspace{-5.5cm}
\centerline{\hspace{ 8.2cm}
\psfig{figure=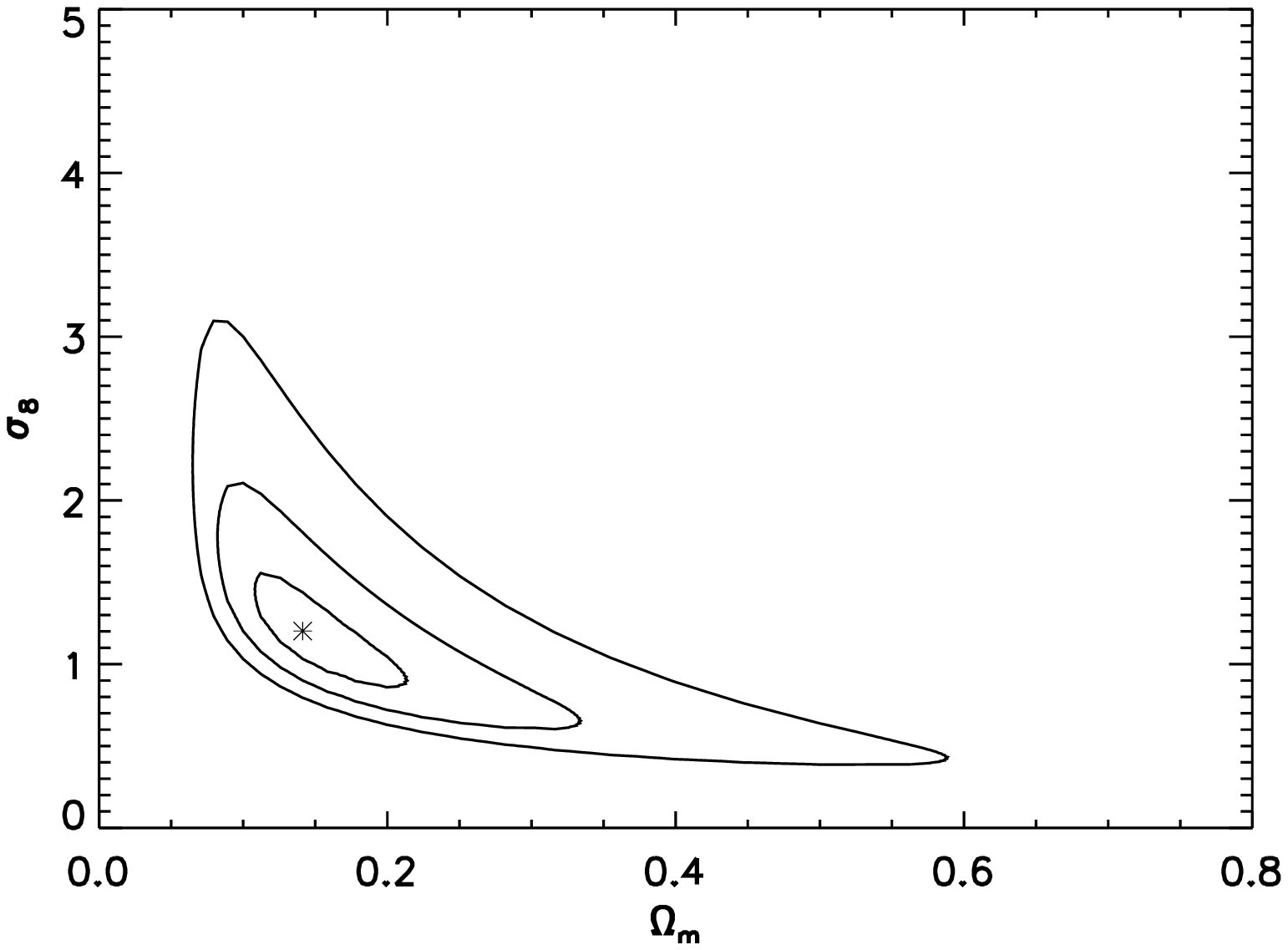,height=5.5cm,width=9.0cm}}
\vspace{-0.35cm}
\caption{\small Likelihood contours (68.3, 95.4, 99.0\%) of the REFLEX
sample in the $P_0$-$\Omega_m$ (left panels) and in the
$\sigma_8$-$\Omega_m$ parameter space (right panels). The $\sigma_8$
values are computed with the high-peak biasing (Kaiser 1984). The
amplitudes, $P_0$, of the power spectra are given in units of
$h^{-4}\,{\rm Mpc}^4$.  The upper row show the results for the 342
REFLEX clusters located within comoving distances $r\le
500\,h^{-1}\,{\rm Mpc}^{-1}$ ($z\le 0.18$) for the fiducial cosmology.
The lower row shows the results for the 403 clusters within $r\le
750\,h^{-1}\,{\rm Mpc}$ ($z\le 0.27$) for the same fiducial
cosmology. The crosses mark the points with the highest likelihood
value.}
\label{FIG_LOG1}
\end{figure*}
\begin{figure*}
\vspace{-0.0cm}
\centerline{\hspace{-10.0cm}
\psfig{figure=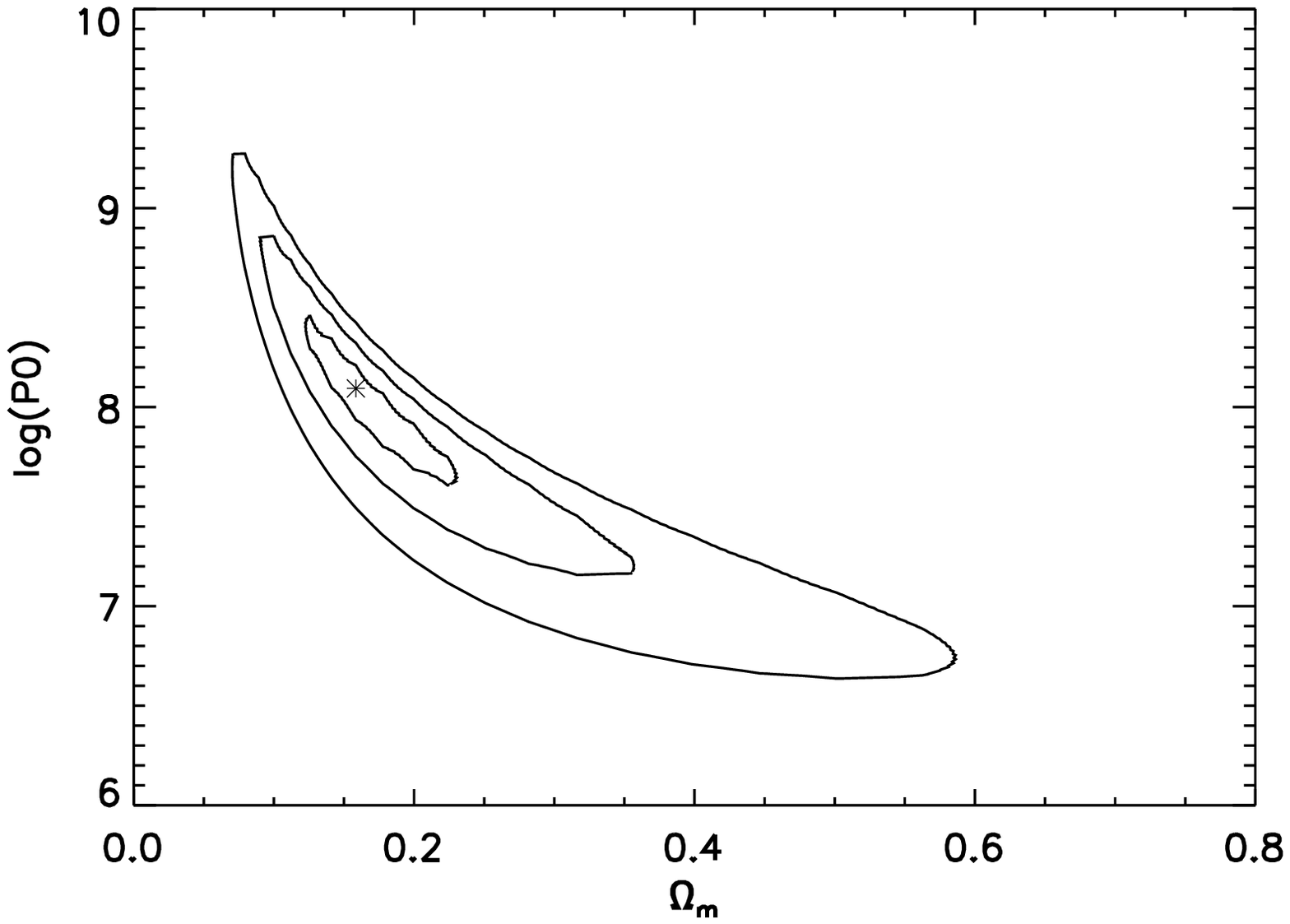,height=5.5cm,width=9.0cm}}
\vspace{-5.5cm}
\centerline{\hspace{ 8.2cm}
\psfig{figure=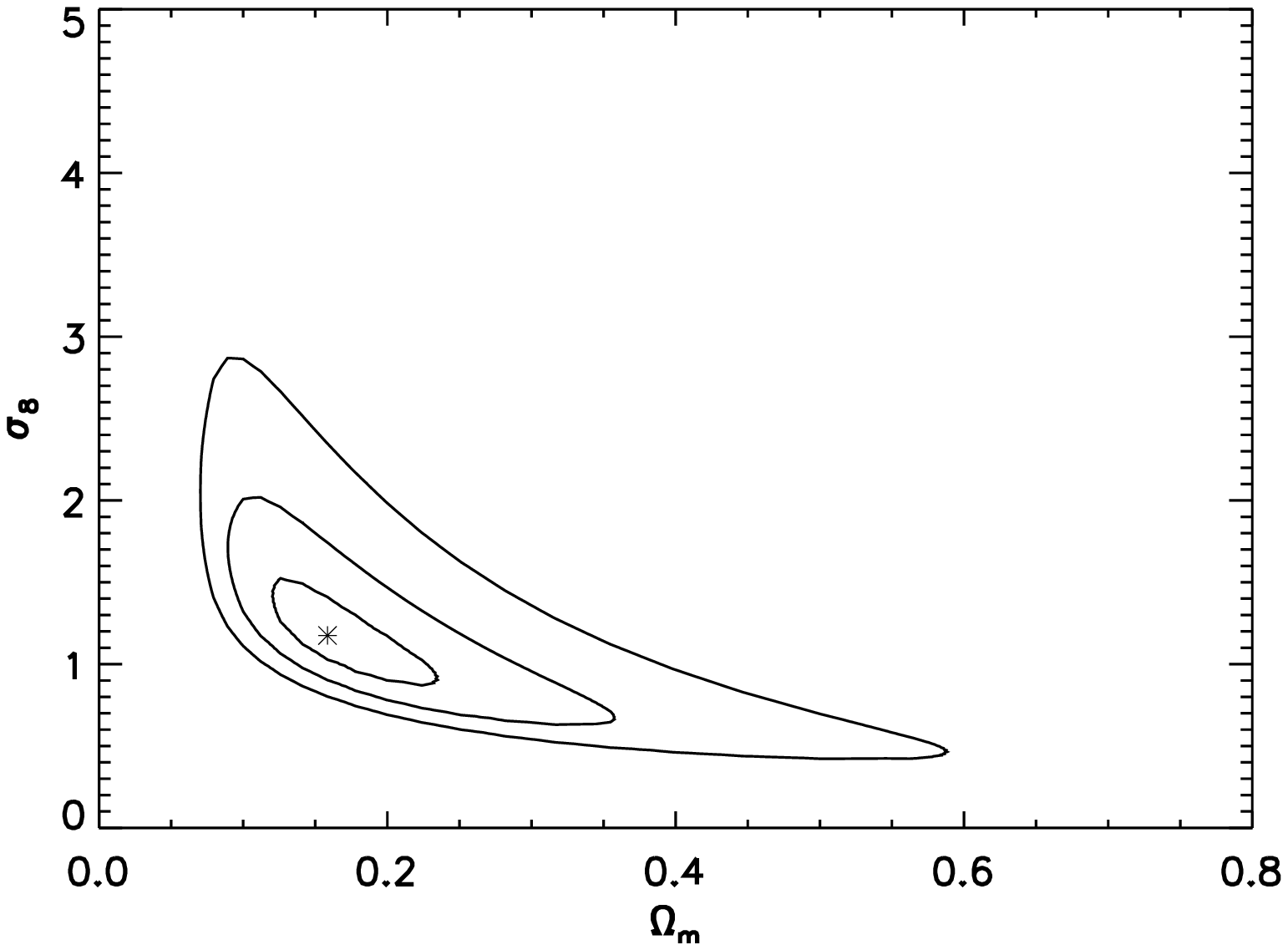,height=5.5cm,width=9.0cm}}
\vspace{-0.5cm}
\centerline{\hspace{-10.0cm}
\psfig{figure=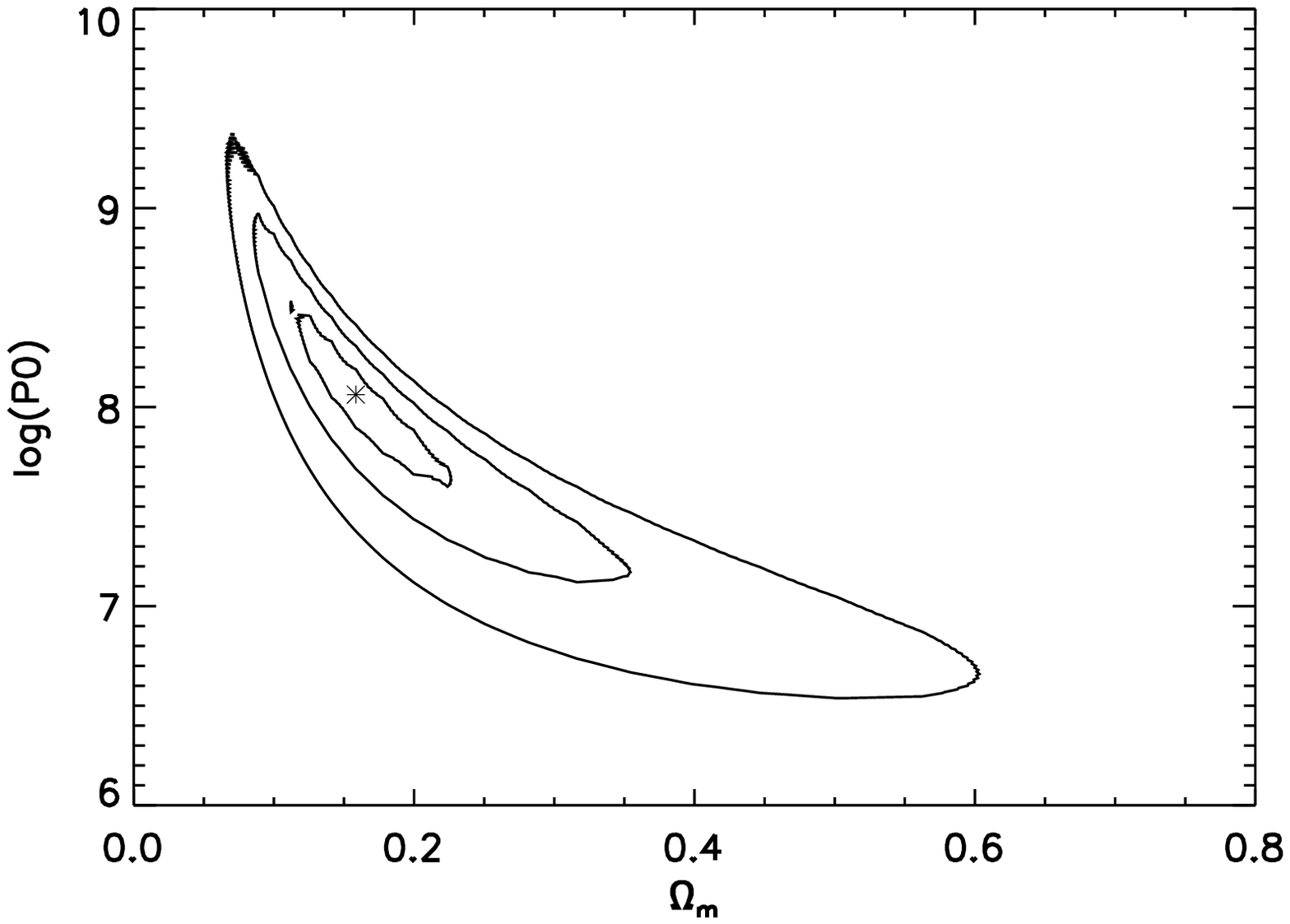,height=5.5cm,width=9.0cm}}
\vspace{-5.5cm}
\centerline{\hspace{ 8.2cm}
\psfig{figure=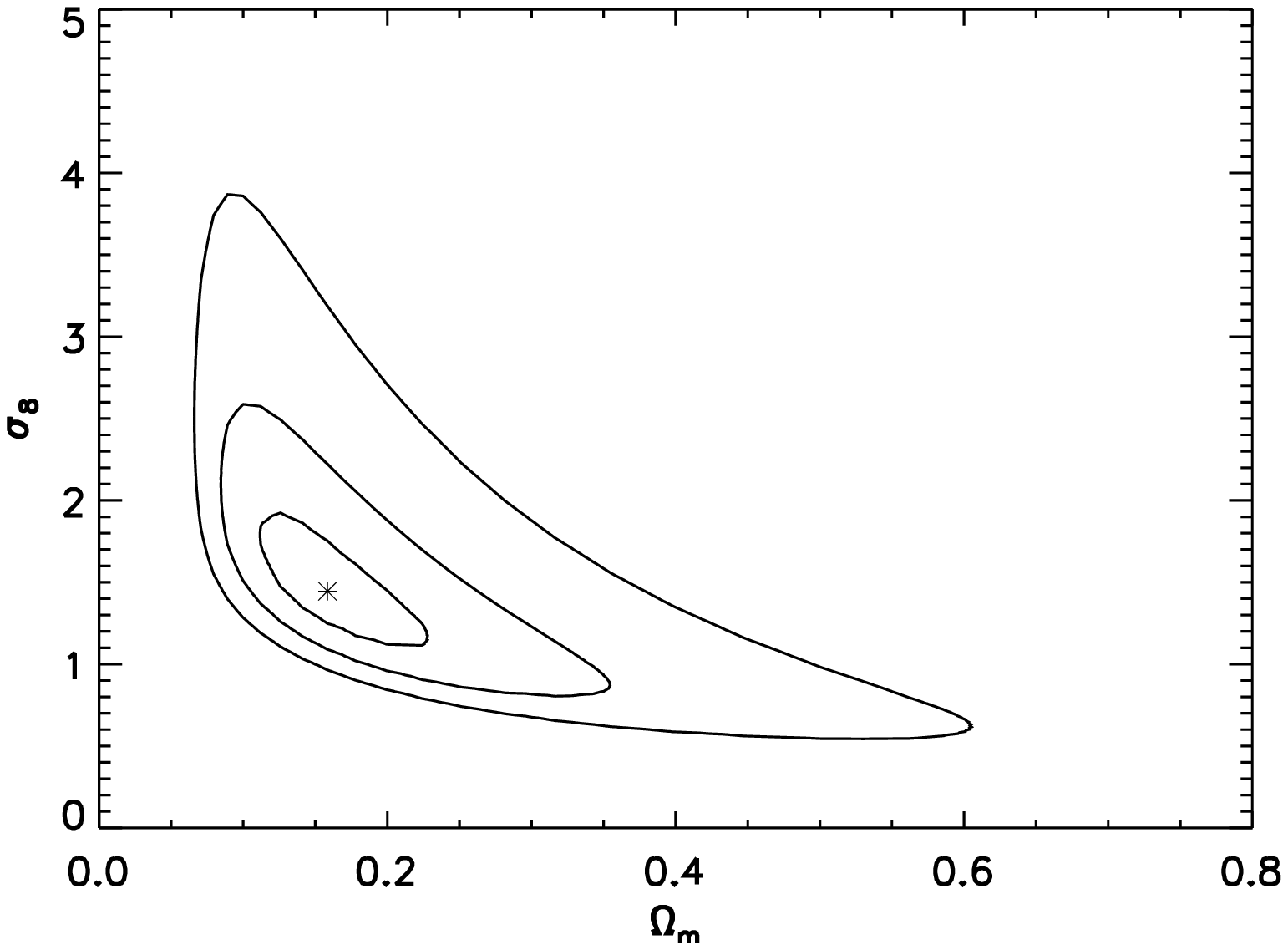,height=5.5cm,width=9.0cm}}
\vspace{-0.35cm}
\caption{\small Upper row: Likelihood contours as in the upper row of
Fig.\,\ref{FIG_LOG1} but for the Einstein-de Sitter fiducial
cosmology. Lower row: Likelihood contours using the empirical
mass/X-ray luminosity relation of Reiprich \& B\"ohringer (2002, see
eq.\,\ref{ML}), but with cluster masses artificially boosted by a
factor of two with respect to the observed relation.}
\label{FIG_LOG2}
\end{figure*}

\begin{figure*}
\vspace{-0.0cm}
\centerline{\hspace{-10.0cm}
\psfig{figure=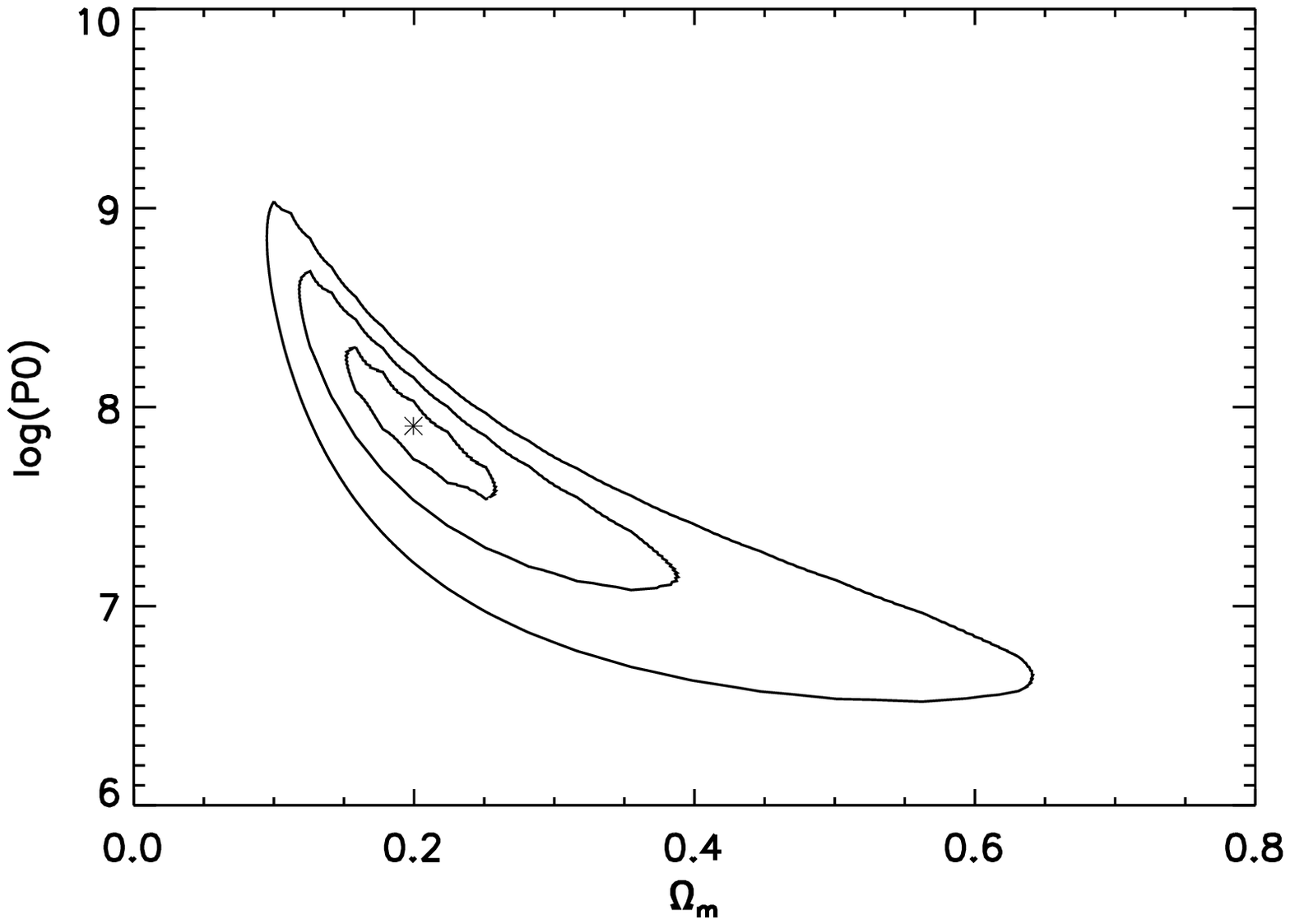,height=5.5cm,width=9.0cm}}
\vspace{-5.5cm}
\centerline{\hspace{ 8.2cm}
\psfig{figure=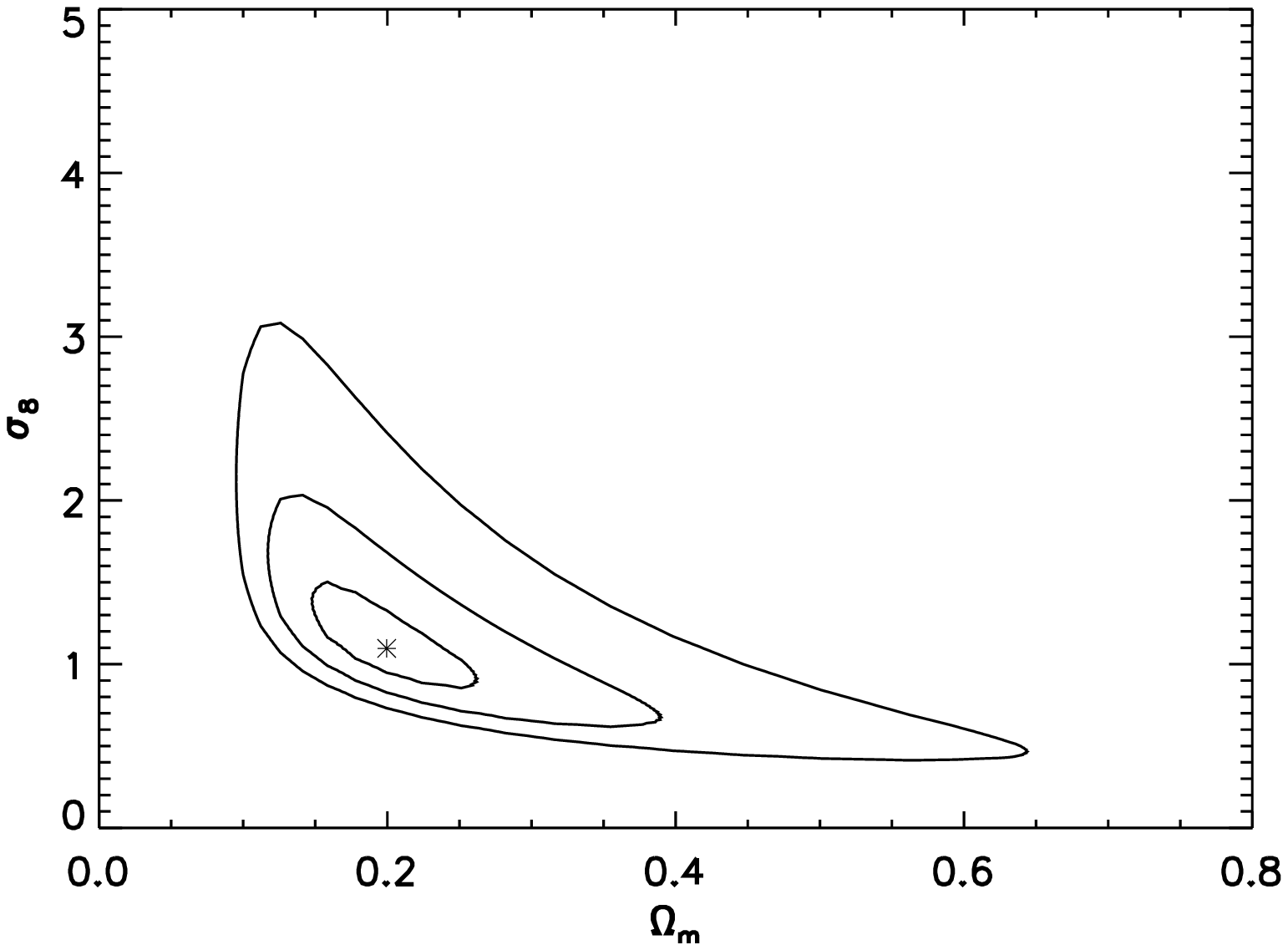,height=5.5cm,width=9.0cm}}
\vspace{-0.5cm}
\centerline{\hspace{-10.0cm}
\psfig{figure=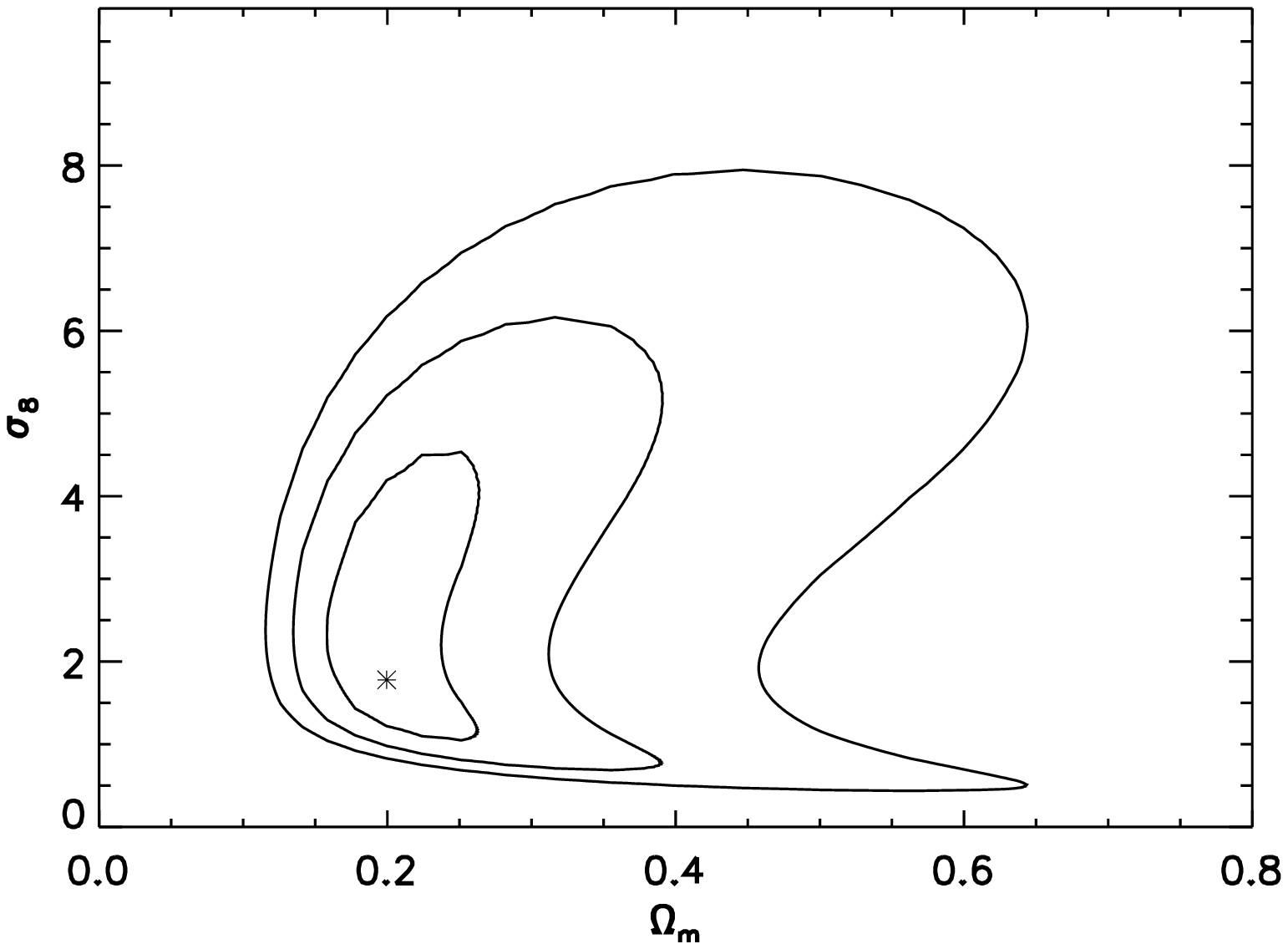,height=5.5cm,width=9.0cm}}
\vspace{-5.5cm}
\centerline{\hspace{ 8.2cm}
\psfig{figure=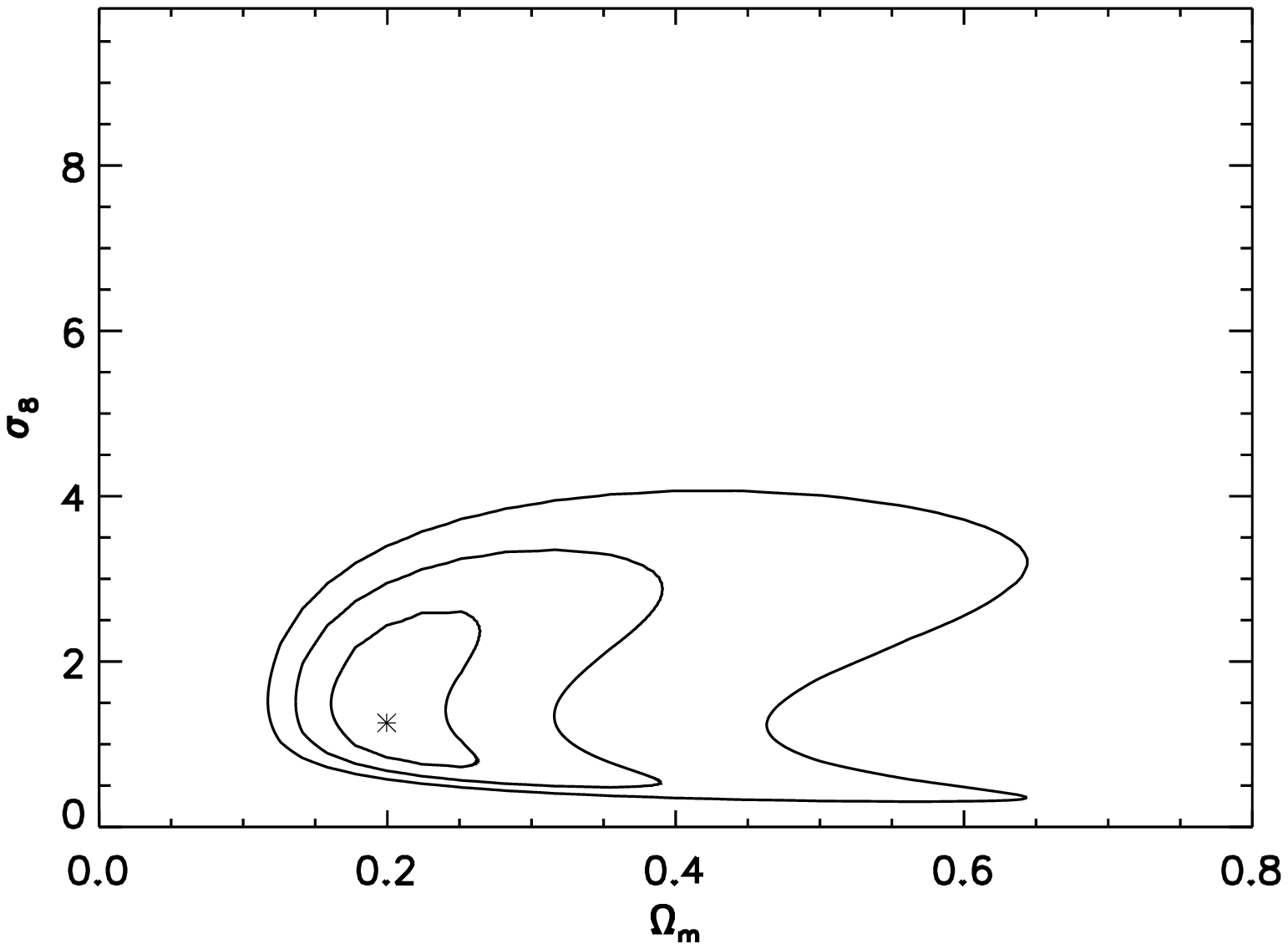,height=5.5cm,width=9.0cm}}
\vspace{-0.35cm}
\caption{\small Likelihood contours for the fiducial cosmology but
with the $2\sigma$ upper limit of the baryon density,
$\Omega_bh^2=0.029$, measured with the BOOMERanG experiment (Masi et
al. 2002) plotted in the $P_0$-$\Omega_m$ parameter space (upper left)
and for the biasing models of Kaiser (1984, upper right), Mo \& White
(1996, lower left) and Sheth \& Tormen (1999, lower right). Note the
different scalings of the $\sigma_8$ axes.}
\label{FIG_LOG3}
\end{figure*}

\section{The REFLEX sample}\label{REFLEX}

The REFLEX sample has 452 southern X-ray clusters of galaxies, 449
with measured redshifts, $z\le 0.45$ (B\"ohringer et al. 2001). The
clusters are selected in an area of 13\,924 square degrees (4.24\,sr)
from the ROSAT All-Sky Survey (RASS, Tr\"umper 1993, Voges et
al. 1999). The nominal limit of the unabsorbed X-ray fluxes is
$3\,10^{-12}\,{\rm erg}\,{\rm s}^{-1}\,{\rm cm}^{-2}$ in the energy
range $(0.1-2.4)$\,keV. 65 percent of the sample are
Abell/ACO/Supplement clusters.

In order to reduce strong spatial variations of the sampling, the 452
REFLEX clusters were selected outside the galactic plane (galactic
latitudes $|b|>20$\,deg) and some additional crowded stellar fields
(e.g., Magellanic Clouds). The remaining corrections for the satellite
exposure time and galactic absorption are well-documented and can be
modelled in detail (e.g., B\"ohringer et al. 2001). The sample has
been successfully used for the determination of the X-ray luminosity
function (B\"ohringer et al. 2002), for the analyses of the cluster
correlation function (Collins et al. 2000), the related peculiar
motions (L. Guzzo et al., in preparation), and the power spectrum
(Schuecker et al. 2001).

Several incompleteness tests described in these REFLEX papers are
based on either the REFLEX sample itself or other observed or
simulated cluster samples. The tests suggest the absence of a
significant incompleteness for clusters with X-ray luminosities
$L_{\rm X}\ge 2.5\,10^{42}\,h^{-2}\,{\rm erg}\,{\rm s}^{-1}$.  The
present investigation uses the 428 clusters which have at least 10
X-ray source counts and which fall within this well-controlled
luminosity range. For comoving distances $r\le 500\,h^{-1}\,{\rm Mpc}$
($z\le 0.18$) no systematic deficiencies of the comoving REFLEX
cluster number densities are found.

\section{The REFLEX KL eigenvectors and eigenvalues}\label{EIGENMODES}

A spherical volume containing the REFLEX survey up to a certain
maximum comoving radius, $r$, is devided into 1\,000 volume elements
(spherical coordinates): 10 angular bins in Right Ascension, 10 in
Declination, and 10 bins along the comoving radial axis. The numbers
of REFLEX clusters and random sample points (see Sect.\,\ref{MEAN}) in
each of the cells are counted, and standard linear algebra codes
(Press et al. 1989) are used to compute the eigenvectors and
eigenvalues of the pixel-averaged whitened correlation matrix. The
following KL analysis is restricted to the $M=540$ eigenvectors with
nonzero eigenvalues, sorted (ranked) with decreasing $<B_n^2>$, i.e.,
with decreasing signal-to-noise. A few examples of one-dimensional
tracings of the three-dimensional eigenvectors are shown for the
largest eigenvalues in Fig.\,\ref{FIG_EV}.

The spectrum of the REFLEX KL eigenvalues is shown in
Fig.\,\ref{FIG_EW_RANK}. The spectrum basically follows a power
law. This indicates that, excluding the extreme $n$ ranges, the KL
eigenvectors sample three-dimensional structures over a large (but not
the complete) $n$ range.

The frequency distribution of the normalized KL eigenvalues gives
information about the Gaussianity of the discrete fluctuation field
and is thus quite important for the justification of the multivariate
Gaussian likelihood functions which will be used for the estimation of
the power spectrum parameters (see Sect.\,\ref{EIGEN_MOD}). The
histogram of the normalized deviations shown in Fig.\,\ref{FIG_DB_N}
is consistent with a Gaussian random distribution on the 93.4\%
confidence level (KS test) and thus supports our basic
assumption. Note that this result is mainly determined with cells
larger than $(50\,h^{-1}\,{\rm Mpc})^3$. For smaller cells deviations
from Gaussianity are expected and other likelihood functions must be
used. The linearity of the KL transform suggests that the Gaussian
distribution of the KL coefficients translates into a Gaussian random
field of the underlying matter distribution. This favours the biasing
model proposed by Kaiser (1984, see Sect.\,\ref{HP_BIAS}).

\section{Results}\label{RESULTS}

The KL method was tested with 27 independent mock cluster samples
selected from the Hubble Volume Simulation. The simulations have the
same fiducial cosmology as used here, $\Omega_m=0.3$ and
$\sigma_8=0.9$, including the values of $n$, $h$, $T_{\rm CMB}$,
$\Omega_bh^2$ mentioned in Sect.\,\ref{TRANS}. The details are given
in Appendix \ref{TESTS}. For each cluster sample and biasing model,
$\Omega_m$ and $\sigma_8$ are varied within suitable intervals, and
the resulting model covariance matrixes (\ref{ML_3}) are computed. The
maximum of the likelihood (\ref{ML_4}) is used to select the best
estimate of $\Omega_m$ and $\sigma_8$ for each sample and biasing
model. The sample-to-sample variations of the parameter values give at
least for the fiducial cosmology an estimate of the errors of
$\Omega_m$ and $\sigma_8$ when the effects of cosmic variance are
included.

The mean and $1\sigma$ errors as obtained from the simulations are for
the matter density $\Omega_m=0.28\pm0.14$ and for the linear matter
normalization $\sigma_8=0.87\pm0.32$ (high-peak biasing),
$\sigma_8=1.20\pm0.66$ (Mo \& White biasing), $0.82\pm0.43$ (Sheth \&
Tormen biasing). Note that for the computation of the mean and
standard deviation of the latter two biasing models only the
likelihood maximum which is located in the high biasing regime was
used (see Sect.\,\ref{O_BIAS}). Below we will compare these errors
with the errors provided by the likelihood contours computed with the
KL method. Compared to the input values of the Hubble Volume
Simulation, no significant systematic errors are thus found (see also
Fig.\,\ref{FIG_HISTOSIM} in Appendix \ref{TESTS}).

The KL method was then applied to the REFLEX cluster sample. The main
results are plotted in the upper panels of Fig.\,\ref{FIG_LOG1}. Shown
are the $1$-$3\sigma$ likelihood contours in the $P_0$-$\Omega_m$ and
$\sigma_8$-$\Omega_m$ parameter spaces. The cosmic matter density with
the highest likelihood value is $\Omega_m=0.16\pm 0.06$ ($1\sigma$
error without cosmic variance and no marginalization with respect to
$h$). It will be seen that the $\Omega_m$ values are basically
unaffected by the assumed biasing model used to compute $\sigma_8$.

The $\sigma_8$ values shown in the right panel of Fig.\,\ref{FIG_LOG1}
are determined with the biasing scheme of Kaiser (1984). The results
obtained with the Mo \& White (1996) and Sheth \& Tormen (1999)
biasing models are discussed below (see Fig.\,\ref{FIG_LOG3}). The
linear normalization of the matter power spectrum with the highest
likelihood value is $\sigma_8=1.2\pm 0.3$ ($1\sigma$ error without
cosmic variance and no marginalization with respect to $h$).

The sensitivity of the results on several given parameter values is
illustrated in the lower panels of Fig.\,\ref{FIG_LOG1} and in
Figs.\,\ref{FIG_LOG2} to \ref{FIG_LOG3}.

In Fig.\,\ref{FIG_LOG1} the results obtained within a maximum comoving
distance of $r=500\,h^{-1}\,{\rm Mpc}$ corresponding to $z=0.18$
(upper panels) are compared to the results obtained within
$r=750\,h^{-1}\,{\rm Mpc}$ or $z=0.27$ (lower panels).  It is seen
that not much information on the spatial fluctuations is gained by the
KL method when the REFLEX clusters outside the well-tested redshift
range (see Sect.\,\ref{REFLEX}) are included.

The upper panels of Fig.\,\ref{FIG_LOG2} show the likelihood contours
determined with an Einstein-de Sitter fiducial cosmology. Due to the
fact that the large-scale structures are mainly probed at $z<0.18$,
the effects of different fiducial cosmologies are not very large and
do not really modify the present KL results.

In the lower panels of Fig.\,\ref{FIG_LOG2} the KL results are shown
where we used the empirical mass/X-ray luminosity relation of Reiprich
\& B\"ohringer (2002), but with a systematic shift applied towards
larger X-ray masses by a factor of two, or equivalently a shift
towards smaller X-ray luminosities by a factor of 2.5. The shift
should ``compensate'' for several possible sources of systematic
errors which could modify the empirical relation (e.g., underestimated
X-ray masses, contamination of the X-ray flux by active galactic
nuclei, relations derived from flux-limited samples). The KL results
show that even large changes in the mass/luminosity relation in the
given directions do not affect the estimation of the matter
density. In the present case, only the normalization of the matter
power spectrum is increased by 20\%.

In the upper panels of Fig.\,\ref{FIG_LOG3} we show the results
obtained with the fiducial cosmology and the $2\sigma$ upper limit on
the baryon density of $\Omega_bh^2=0.029$ obtained from the
combination of BOOMERanG and COBE/DMR data (Masi et al. 2002). The
main effect of the baryons is to steepen $P(k)$. A large $\Omega_b$
value can thus be compensated by a large $\Omega_m$ as seen in
Fig.\,\ref{FIG_LOG3}. The same effect is found in the 2dF 100k data
(Percival et al. 2001, Tegmark, Hamilton \& Xu 2001).

In Fig.\,\ref{FIG_LOG3} we show the likelihood contours determined
with all three biasing models described in Sect.\,\ref{RELATION}. As
mentioned above, $\Omega_m$ is mainly independent of the assumed
biasing model, but an effect is seen in the derived $\sigma_8$ values
which will be discussed below in more detail.

The most important cosmological constraint derived from the spatial
fluctuations of the REFLEX clusters is the cosmic matter density
obtained from the marginalization of the likelihood distributions
shown in Figs.\,\ref{FIG_LOG1} to \ref{FIG_LOG3}. For $h=0.7$ the
REFLEX data give the 95.4\% confidence interval
\begin{equation}\label{OMEGA}
0.07<\Omega_m<0.38\,\,(95.4\%\,\,{\rm
without\,\,cosmic\,\,variance})\,.
\end{equation}
Note that for $\Omega_bh^2=0.029$ the highest likelihood value of most
models is at $\Omega_m=0.20$. A more systematic analysis of models
with different $\Omega_b$ values (and $n$) is necessary and will be
given in the next paper.

The KL analysis of the spatial fluctuations of the REFLEX clusters is
less sensitive to the linear $\sigma_8$. From the marginalization of
the likelihood distributions and for the high-peak biasing model of
Kaiser (1984) we obtained
\begin{equation}\label{SIGMA8}
0.6<\sigma_8<2.6\,\,(95.4\%\,\,{\rm without\,\,cosmic\,\,variance})\,,
\end{equation}
with the highest likelihood value at $\sigma_8=1.2$. 

For the biasing schemes of Mo \& White (1996) and Sheth \& Tormen
(1999) the situation is more complex. In the error-free case one would
expect for the two biasing models two well-separated likelihood
regions centered on the same $\Omega_m$ but at two different
$\sigma_8$ values (see Sect.\,\ref{O_BIAS}). However, the
comparatively large statistical scatter of the observed $P_0$ values
smeares out the high- and low-biasing regimes and thus leads to the
'shoe-like' contours seen in the lower panels of Fig.\,\ref{FIG_LOG3}.
Nevertheless, the results obtained with the simulations (see Appendix
\ref{TESTS}) shows that all three biasing schemes give similar
results, when for the Mo \& White and Sheth \& Tormen models the
$\sigma_8$ values located in the high-biasing regime are selected.

The comparison of the errors of the parameter values obtained from the
REFLEX data and from the simulations which include cosmic variance
indicates that the errors including cosmic variance are about 50\%
larger compared to the KL errors.

\begin{table}
{\bf Tab.\,1.} Comparison of the 95.4\% confidence ranges for
$\Omega_mh^2$ obtained with galaxy clusters (REFLEX), recent
measurements of CBM temperature fluctuations (BOOMERanG, DASI) and
with galaxies (SDSS, 2dFGRS). References: (1) this work, (2)
Netterfield et al. (2001), (3) Szalay et al. (2001), (4) Pryke et
al. (2001), (5) Parcival et al. (2001). SDSS measures the shape
parameter, $\Gamma$, 2dFGRS measures $\Omega_mh$. These values are
transformed using $h=0.7$ and $\Omega_bh^2=0.0196$. REFLEX, SDSS, and
2dFGRS assume a flat universe with a cosmological constant. BOOMERanG
and DASI results have the weakest priors.\\
\vspace{-0.0cm}
\begin{center}
\begin{tabular}{llcc}
Data       &Probe     & $\Omega_mh^2$ & Ref.\\
\hline
REFLEX     &Clusters  & $0.03-0.19$   & (1)\\
BOOMERanG  &CMB       & $0.05-0.25$   & (2)\\
SDSS       &Galaxies  & $0.08-0.20$   & (3)\\
DASI       &CMB       & $0.08-0.24$   & (4)\\
2dFGRS     &Galaxies  & $0.10-0.18$   & (5)\\
\hline
\end{tabular}
\end{center}
\end{table}

\section{Discussion}\label{DISCUSS}

The present investigation applies the KL method to estimate the values
of the cosmic matter density and the linear normalization of the
matter power spectrum. The fluctuations of the comoving densities of
the REFLEX clusters are analyzed up to Gpc scales with a well-defined
survey specific set of eigenvectors. This offers the possibility to
analyse the fluctuations up to Gpc scales without the disturbing
effects of correlations between different power spectral densities,
$P_{\rm obs}(k)$, which affects all previous cluster measurements on
the largest scales. Note that the correlations artificially reduce the
statistical errors, so that simple numerical model fits to $P_{\rm
obs}(k)$ in order to estimate the values of the cosmological
parameters cannot be applied.

The main result obtained with the KL analysis of the REFLEX clusters
is that for spatially flat CDM-like structure formation scenarios the
data support a low-density universe with (rescaling $\Omega_m$ to
$\Omega_mh^2$ using our prior $h=0.7$)
\begin{equation}\label{RES}
0.03<\Omega_mh^2<0.19\,,
\end{equation}
and the linear normalization $0.6<\sigma_8<2.6$ (95.4\% confidence
intervals without cosmic variance). The notation underlines the fact
that we did not marginalize the results for different $h$ values. The
Einstein-de Sitter case is ruled out with 99.99\% confidence. The
errors obtained with the KL method include marginalization over
several important reduction parameters but not cosmic variance. We
have estimated the effect using 27 REFLEX-like mock samples selected
from the Hubble Volume Simulation, and found that for the current
sample the KL errors are probably underestimated by 50\%.

We want to stress that the $\Omega_m$ measurements appear to be quite
robust against several partially quite drastic changes of important
reduction parameters. What really matters seems to be the baryon
density (and thus also the spectral index, $n$, of the primordial
power spectrum). A systematic study of models with different
$\Omega_b$ and $n$ is in preparation.

The REFLEX confidence range for $\Omega_mh^2$ in (\ref{RES}) is in
good agreement with other recent measurements (see Tab.\,1). The table
gives the 95.4\% confidence intervals obtained from different
measurements with the minimum number of priors (and not results
obtained by combined data sets). Note that the different groups
measured $\Gamma$, $\Omega_m$, $\Omega_mh$, or $\Omega_mh^2$, and give
either $1\sigma$ or $2\sigma$ errors. We have tried to transform the
original results to $\Omega_mh^2$ and 95.4\% errors, having in mind
that this can only be done approximately.

The Sloan Digital Sky Survey (SDSS) result is obtained from the galaxy
clustering of 222 square degrees early imaging data (Szalay et
al. 2001). For SDSS the shape parameter, $\Gamma$, of the power
spectrum is transformed to $\Omega_mh^2$ assuming $h=0.7$ and
$\Omega_bh^2=0.0196$ and the approximate formula given in Sugiyama
(1995). The 2dF Galaxy Redshift Survey (2dFGRS) result for $\Omega_mh$
is obtained with 166\,490 galaxies. The Degree Angular Scale
Interferometer (DASI) and BOOMERanG experiments measure the angular
power spectrum of the CMB anisotropy. REFLEX, SDSS, and 2dFGRS assume
a flat universe with a cosmological constant, but the results do not
strongly depend on $\Omega_\Lambda$. They also assume $n=1$. REFLEX
has the additional constraint $\Omega_bh^2=0.0196$. The BOOMERanG
results have the weak prior $0.45<h<0.90$ and eliminates models where
the Universe is younger than 10\,Gyr. DASI assumes $h>0.45$ and the
optical depth due to reionization $0.0\le \tau_c\le 0.4$. Compared to
the results shown in Tab.\,1, the REFLEX results extents to slightly
smaller $\Omega_mh^2$ values. Smaller confidence ranges from REFLEX
are expected when the KL analysis will include both the fluctuations
and the mean cluster number densities, utilizing the complementarity
of clustering and abundance of clusters. In this way the KL analysis
will allow us to fully exploit the cosmological potential of the
REFLEX survey of X-ray clusters.

The `banana-shape' likelihood contours obtained with the REFLEX data
(see Figs.\,\ref{FIG_LOG1} to \ref{FIG_LOG3}) might be taken as an
indication that the primordial power spectrum is less constrained by
the current REFLEX data. A significant improvement is expected when
the southern REFLEX sample and the northern NORAS sample (B\"ohringer
et al. 2000) are extended to the deeper flux limit of
$2\,10^{-12}\,{\rm erg}\,{\rm s}^{-1}\,{\rm cm}^{-2}$ in the energy
range 0.1-2.4\,keV and combined to an all-sky sample of about 1\,700
X-ray selected clusters of galaxies.

We would like to thank the REFLEX group for their help in the
preparation of the X-ray cluster sample, D. Eisenstein and W. Hu for
the computer code for the matter transfer functions, the Virgo
Consortium for the simulated LCDM cluster sample, and the referee
Stefano Borgani for his useful comments. P.S. acknowledges
support under the grant No.\,50\,OR\,9708\,35.

\appendix

\section{Validation of the KL estimation of the power spectrum parameters
with N-body simulations}\label{TESTS}

Mock samples are used to test the likelihood method, especially
systematic errors and the effects of cosmic variance. For studies of
the clustering properties of X-ray selected cluster samples a crucial
step is the transformation of the simulated cluster gravitational
masses to the observable X-ray luminosities. Here we use the empirical
mass/X-ray luminosity relation of Reiprich \& B\"ohringer (2002) for
the energy range (0.1-2.4)\,keV,
\begin{equation}\label{ML}
\frac{L_{\rm X}}{h^{-2}10^{44}\,{\rm erg}\,{\rm s}^{-1}}\,=\,7.199\times
10^{-20}\,\left(\frac{M}{h^{-1}M_\odot}\right)^{1.31}\,,
\end{equation}
assuming a negligible intrinsic scatter. The formal $1\sigma$ errors
of the scaling factor and the index of (\ref{ML}) are 6.3 and 5.1\%,
respectively. In order to apply this relation, one has to ensure that
the cluster masses as defined through the simulations are consistent
with the masses of the empirical mass/X-ray luminosity relation as
defined through the X-ray measurements. For the present error
estimation we use the simple mass transformation model described below
giving redshift histograms similar to REFLEX. For exact model
comparisons going beyond simple error estimates, more refined
transformation models or simulations adapted to the empirical
mass-luminosity relation should be used.

\subsection{Simulated clusters}\label{SIMUL}

The Virgo Consortium provides cluster catalogues extracted from the
Hubble Volume Simulations (see, e.g., Jenkins et al. 2001, see also
Evrard et al. 2001). The public cluster catalogue used here is
selected at $z=0$ from one $\Lambda$Cold Dark Matter (LCDM) simulation
with a box length of $3\,h^{-1}\,{\rm Gpc}$ and $\Omega_m=0.3$,
$\Omega_\Lambda=0.7$, and $\sigma_8=0.90$. The LCDM transfer function
was computed with the CMBFAST routine (Seljak \& Zaldarriaga 1996)
assuming $h=0.7$, $\Omega_b h^2=0.0196$ (Burles \& Tytler 1998), and a
primordial slope of $P(k)$ of unity. Each of the $10^9$ particles has
a mass of $2.22\,10^{12}\,h^{-1}\,M_\odot$. Motivated by the spherical
collaps model the Virgo Consortium attempts to identify virialized
regions that are overdense by a factor $\sim 200$ applying the
friend-of-friend group finder with a linking length of 0.164. The
resulting catalogue comprises 1\,560\,995 clusters.  The minimum
number of particles per cluster is 30.

The friend-of-friend cluster masses obtained from the simulations are
measured out to a radius, $r_{\rm sim}$, where the averaged density
contrast relative to the {\it local} mass density is approximately
324. For the empirical mass/X-ray luminosity relation, Reiprich \&
B\"ohringer used the radius $r_{200}$, where the average density
contrast of 200 is related to the {\it Einstein-de Sitter} critical
mass density. The mass conversion factor is obtained from the relation
between virial mass and X-ray temperature obtained from hydrodynamical
simulations (e.g., Bryan \& Norman 1998), giving for fixed temperature
and redshift $M(r_{200})/M(r_{\rm sim})=0.69$. A slightly better match
between simulated and observed redshift histograms is obtained with
the conversion factor 0.67 which we finally used.

The $M(r_{200})$ cluster masses are transformed to X-ray luminosities
in the energy range (0.1-2.4\,keV). The observer restframe fluxes are
obtained with the cluster luminosity distance taking into account the
cosmic K-correction as obtained with a refined Raymond-Smith code (the
cluster X-ray temperatures are estimated with the $L_{\rm X}$-$T$
relation of Markevitch (1998) without cooling flow corrections). The
resulting total fluxes are reduced by 10 percent to get the measured
fluxes because the X-ray observations do not include the flux in the
outer wings of the cluster X-ray image. This average difference
between total and observed fluxes is obtained from Monte-Carlo
simulations (H. B\"ohringer, in preparation). The variation of the
X-ray flux limit of the REFLEX sample across the survey area are
computed in the same way as in Schuecker et al. (2001).

\subsection{Comparison of true and estimated parameter values}\label{TRUE}

\begin{figure}
\vspace{-0.0cm}
\centerline{\hspace{-0.3cm}
\psfig{figure=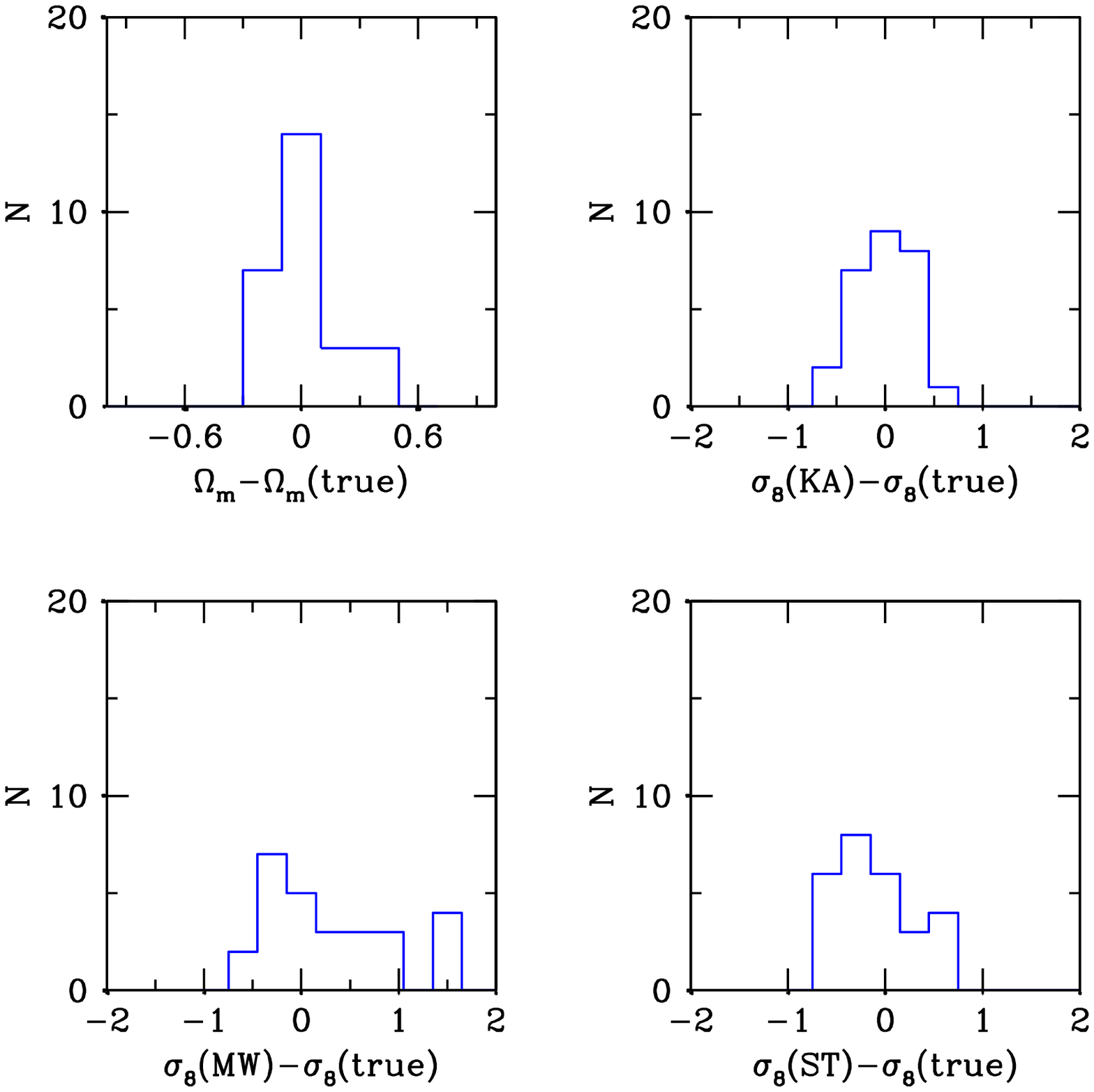,height=9.5cm,width=9.5cm}}
\vspace{-1.20cm}
\caption{\small Histograms of the differences between the KL estimate
of matter density, $\Omega_m$, and the linear normalization of the
matter power spectrum, $\sigma_8$, with the input (true) values of the
simulations. The biasing schemes are denoted by KA (high-peak biasing
of Kaiser 1984), MW (Mo \& White 1996) and ST (Sheth \& Tormen
1998). The frequency distributions are obtained with 27 REFLEX-like
subsamples selected from the Hubble Volume Simulation.}
\label{FIG_HISTOSIM}
\end{figure}

We selected 27 independent REFLEX-like subsamples from the LCDM Hubble
Volume cluster sample. The average number of clusters per sample and
its standard deviation is $435\pm 28$, similar to the 428 REFLEX
clusters used for the final analyses. The redshift histograms closely
resemble the observed distribution. We thus expect realistic error
estimates from the simulations.

The histograms of the residuals between the values estimated with the
KL method and the true (simulation input) values for $\Omega_m$ and
for $\sigma_8$ as obtained with the three biasing models (see
Sect.\,\ref{RELATION}) are shown in Fig.\,\ref{FIG_HISTOSIM}.

The frequency distribution of the KL estimates of $\Omega_m$ gives the
formal mean and standard deviation of $\Omega_m=0.32\pm0.19$ (see
Fig.\,\ref{FIG_HISTOSIM}, upper left).  Note that the distribution is
slightly skewed, the median value is $\Omega_m=0.27$.  A $2\sigma$
clipping rejects two measurements and gives $\Omega_m=0.28\pm
0.14$. Whereas the mean value turns out to be quite stable, it appears
to be more difficult to get a stable estimate of the error which
includes cosmic variance. A still larger number of simulations is
necessary to improve the accuracy. For the comparison with the
internal errors given by the KL method (see Sect.\,\ref{RESULTS}) we
use the latter more stable estimate, keeping in mind that the error
could be underestimated by about 30\%.

For the linear matter normalization the following formal means and
standard deviations are obtained: $\sigma_8=0.87\pm0.32$ (high-peak
biasing, Fig.\,\ref{FIG_HISTOSIM} upper right), $\sigma_8=1.20\pm0.66$
(Mo \& White biasing, Fig.\,\ref{FIG_HISTOSIM} lower left),
$0.82\pm0.43$ (Sheth \& Tormen biasing, Fig.\,\ref{FIG_HISTOSIM} lower
right). Note that for the computation of the means and standard
deviations of the latter two biasing models only the values located in
the high biasing regime are used (see Sect.\,\ref{O_BIAS}). The errors
include cosmic variance.

\end{document}